\documentclass[12pt,preprint,eqsecnum]{aastex}
\usepackage{epsfig}

\newcommand*{\eg}{e.g.,\ }
\newcommand*{\ie}{i.e.,\ }
\newcommand*{\eqref}[1]{(\ref{#1})}

\newcommand*{\myplottwo}[2]{\begin{center}\includegraphics[height=7in,clip=true]{#1}\hfil\includegraphics[height=7in,clip=true]{#2}\end{center}}

\slugcomment{Draft}

\shorttitle{KHI in PPD}
\shortauthors{Barranco}

\begin{document}

\title{Three-Dimensional Simulations of Kelvin-Helmholtz Instability in Settled Dust Layers in Protoplanetary Disks}

\author{Joseph A. Barranco}
\affil{Department of Physics \& Astronomy\\
San Francisco State University\\
1600 Holloway Avenue\\
San Francisco, CA 94132}
\email{barranco@stars.sfsu.edu}

\begin{abstract}
As dust settles in a protoplanetary disk, a vertical shear develops because the dust-rich gas in the midplane orbits at a rate closer to true Keplerian than the slower-moving dust-depleted gas above and below.  A classical analysis (neglecting the Coriolis force and differential rotation) predicts that Kelvin-Helmholtz instability occurs when the Richardson number of the stratified shear flow is below roughly one-quarter.  However, earlier numerical studies showed that the Coriolis force makes layers more unstable, whereas horizontal shear may stabilize the layers.  Simulations with a 3D spectral code were used to investigate these opposing influences on the instability in order to resolve whether such layers can ever reach the dense enough conditions for the onset of gravitational instability.  I confirm that the Coriolis force, in the absence of radial shear, does indeed make dust layers more unstable, however the instability sets in at high spatial wavenumber for thicker layers.  When radial shear is introduced, the onset of instability depends on the amplitude of perturbations: small amplitude perturbations are sheared to high wavenumber where further growth is damped; whereas larger amplitude perturbations grow to magnitudes that disrupt the dust layer.  However, this critical amplitude decreases sharply for thinner, more unstable layers.  In 3D simulations of unstable layers, turbulence mixes the dust and gas, creating thicker, more stable layers.  I find that layers with minimum Richardson numbers in the approximate range 0.2 -- 0.4 are stable in simulations with horizontal shear.
\end{abstract}


\section{INTRODUCTION}

It is a remarkable fact that planets start out as microscopic grains within the protoplanetary disks of gas and dust in orbit around newly-formed protostars, somehow growing roughly $10^{40}$ orders of magnitude in mass in a period no more than $10^7$ years corresponding to disk lifetimes \citep{lissauer93}.  There is no one physical process that can explain growth over this enormous range of sizes: the very smallest grains (micron to millimeter sizes) can grow via collisional agglomeration in which the sticking mechanism is electrostatic in nature; whereas, on the other end of the size spectrum, objects in the kilometer to tens of kilometers regime can grow via gravity-enhanced collisions \citep{beckwith00}.  The least understood stage of growth is how millimeter-size particles grow to kilometer-size; grains in this regime are too large for sticking via electrostatic forces, yet far too small to have any significant self-gravity.  Even more problematic is the fact that particles in this intermediate-size regime are strongly affected by aerodynamic drag of the surrounding gas: meter-size objects, for example, have radial drift speeds on the order of $10^4$~cm/s at 1~AU and thus spiral inward onto the protostar on the timescale of a few hundred years \citep{weidenschilling77}.  Whatever process is responsible for grain growth through this range of sizes must act on timescales faster than this inspiral time if any raw materials are to be available to build protoplanets. 

\citet{goldreich73} and \citet{safronov69} proposed that a very thin, very dense sheet of settled dust in the midplane of the protoplanetary disk might be be gravitationally unstable; the nonlinear evolution would result in the layer clumping-up directly into gravitationally-bound kilometer-size planetesimals on a timescale of order the orbital period.  According to this scenario, there is a direct jump from small particles to kilometer-size planetesimals without growing slowly through the intermediate sizes which have very short orbital decay timescales.   As attractive as this mechanism is for planetesimal formation, a significant obstacle is turbulence which can stir and mix the dust with the gas and prevent the dust layer from settling into a thin enough, dense enough sheet for the gravitational instability to operate \citep{weidenschilling80}.

Even in the absence of any mechanism to drive turbulence, the settling of the dust particles into a thin
sublayer in an initially laminar midplane would create a vertical shear that might be unstable to Kelvin-Helmholtz instability.   Because of a relatively weak outward radial pressure gradient, pure gas in a protoplanetary disk orbits the protostar at a rate slightly slower than true Keplerian: $V_{gas} = V_K(1-\eta)$, where $\eta\sim 10^{-3}$ for typical conditions at 1~AU \citep{adachi76,weidenschilling77}.  Pure dust, in the absence of any gas, would orbit exactly at the Keplerian rate.   One can show that in the limit of perfect dust-gas coupling, the orbital velocity of a mixture of gas and dust is a function of the local dust-to-gas ratio \citep{adachi76}; as dust settles into the midplane, the dust-rich gas in the midplane orbits faster than the dust-depleted gas above and below the midplane.  Two-fluid (gas and dust) numerical simulations by \citet{cuzzi93}, \citet{champney95} and \citet{dobrovolskis99} suggested that turbulent diffusion would prevent settling of grains into thin enough sheets for gravitational instability.

The onset of Kelvin-Helmholtz instability is determined by a competition between the stabilizing effects of stratification and the destabilizing effects of vertical shear.  If there is no rotation and no horizontal shear, then a necessary (but not sufficient) condition for instability is the Richardson number criterion \citep{chandra60,drazin81}:
\begin{equation}\label{E:Richardson_criterion}
\textrm{instability: } \quad Ri(z) \equiv \frac{N^2}{(dU/dz)^2} < Ri_{crit} \sim 0.25 \quad \textrm{for some z,}
\end{equation}
where $N$ is the Brunt-V\"{a}is\"{a}l\"{a} frequency (the frequency of buoyant oscillations of a stably-stratified medium) and $dU/dz$ is the vertical shear.  The critical Richardson number of one-quarter corresponds to a state in which the kinetic energy of the vertical shear is sufficient to lift the heavier gas out of the gravitationally potential well and remix it with the lighter overlying fluid.  Assuming that dust would settle into thinner and thinner layers down to the limit set by the classic Richardson criterion,  \citet{sekiya98}, \citet{sekiya00}, and \citet{youdin02} determined critical vertical quasi-equilibrium dust profiles and investigated the conditions necessary for such layers to be gravitationally unstable.  \citet{garaud04} pursued two-fluid linear calculations (without rotation and radial shear) of dust sedimentation into sheets and found that the Kelvin-Helmholtz instability was excited before gravitational instability unless the global dust-to-gas ratio was greatly enhanced over solar abundance. 
\citet{youdin04} suggested that such enhancements might be attainable because the inward drift speed decreases as particles migrate inward, resulting in ``particle pile-ups.''

It is not at all clear whether the classic Richardson number criterion is appropriate for the Kelvin-Helmholtz instability in protoplanetary disks in which both rotation and radial shear could significantly affect the onset  of instability as well as alter the nonlinear evolution of any instability that develops.
\citet{ishitsu03} pursued a purely linear analysis to study the effect of radial shear on the time evolution of unstable Kelvin-Helmholtz modes and showed that the modes were sheared to higher spatial wavenumber and could eventually be stabilized. On the other hand, \citet{gomez05} investigated the effects of the Coriolis force on the onset of instability, and found that settled layers were unstable at much higher Richardson numbers (corresponding to thicker dust layers) than in the classic case.  However, these simulations were two-dimensional and did not include the radial shear.   

The motivation for this work is to investigate the apparently opposing influences of rotation (destabilizing) and radial shear (stabilizing) on the Kelvin-Helmholtz instability of settled dust layers and resolve whether or not such layers can ever reach the thin enough, dense enough conditions for the onset of gravitational instability.   Apart from the impact of Kelvin-Helmholtz instability on the formation of planetesimals, it is important to determine the vertical distribution of dust in order to correctly interpret observations \citep{brittain05, rettig06, dullemond07}.  The approach we take here is that the dust is perfectly coupled to the gas; in terms of timescales, the friction or stopping time (see \eqref{E:epstein} below) is much shorter than the evolution of Kelvin-Helmholtz instabilities if they are present.  We will not simulate the formation of the dust layer from a well-mixed state, but assume that the layers have already formed with a given profile of the dust-to-gas ratio.  If the layer turns out to be unstable on a fast timescale, it implies that such a layer would never have formed in the first place.  In \S2, we present equations for the hydrodynamic evolution of settled dust layers in the limit of perfect dust-gas coupling.  In \S3, we revisit the cases of  (i) no rotation, no shear, and (ii) rotation, no shear.  In \S4, we present new three-dimensional simulations of the evolution of settled dust layers with both rotation and shear.  Finally, in \S5, we discuss the impact of these results on the planetesimal formation via gravitational instability. 

\section{SINGLE-FLUID EQUATIONS FOR PERFECTLY COUPLED GAS \& DUST IN 3D
CARTESIAN SHEARING BOX}

\subsection{Equilibrium for a Gas Disk}

Consider the time-independent, axisymmetric azimuthal flow $\bar{V}_{\phi}$
of gas around a protostar of mass $M_{\star}$:
\begin{equation}\label{E:azimuthal_flow}
\frac{\bar{V}^2_{\phi}}{r}\mathbf{\hat{r}} = \mathbf{\nabla}\Phi + \frac{1}{\bar{\rho}_g}\mathbf{\nabla}\bar{p}_g,
\end{equation}
where $(r,\phi,z)$ are protostar-centered cylindrical coordinates with corresponding unit vectors
$(\mathbf{\hat{r}},\mathbf{\hat{\phi}},\mathbf{\hat{z}})$, $\Phi =  -GM_{\star}/\sqrt{r^2+z^2}$
is the gravitational potential, $G$ is the gravitational constant, and
$\bar{\rho}_g$ and $\bar{p}_g$ are the equilibrium gas density and pressure.
Protoplanetary disks are thermally cool in the sense that the gas sound speed
$c_s$ is much slower than the Keplerian orbital velocity
$V_K(r)\equiv r\Omega_K(r)\equiv\sqrt{GM_{\star}/r}$.   Hydrostatic balance
implies that the time it takes sound waves to traverse the thickness of the
disk is of order the orbital period; thus, cool disks are geometrically thin
\citep[see][]{frank85}:
\begin{mathletters}
\begin{eqnarray} \label{E:hydrostatic}
c_s & \sim & \Omega_K H_g, \label{E:soundspeed}\\
\delta \equiv c_s/V_K & \sim & H_g/r < 1,\label{E:coolthin}
\end{eqnarray}
\end{mathletters}
\noindent where $H_g$ is the vertical pressure scale height. In cool disks,
the radial component of the protostellar gravity nearly balances the
centrifugal force, but because of the relatively weak outward radial
pressure force, the gas orbits at slightly slower than the Keplerian
velocity \citep{adachi76, weidenschilling77}:
\begin{mathletters}
\begin{eqnarray}\label{E:Keplerian}
\Omega(r,z) & = & \Omega_K(r)\left[1-\eta(r,z)\right], \label{E:omegaK}\\
\eta(r,z)   & = & \frac{-(\partial\bar{p}_g/\partial r)/\bar{\rho}_g}{2GM_{\star}/r^2} + \case{3}{4}\left(\frac{z}{r}\right)^2 + \mathcal{O}(\delta^4), \label{E:eta1}
\end{eqnarray}
\end{mathletters}
where the fractional deviation from Keplerian is $\eta\sim\delta^2\ll 1$.
Depending on the disk model, $\eta$ typically takes values between
$10^{-3}-10^{-2}$ at $r\!=\!1$~AU, and the Mach number for the maximum deviation from Keplerian is of order $\eta V_K/c_s \sim 0.1$



In contrast to a pure gas disk which orbits at a slightly sub-Keplerian
speed, a disk of pure (``pressureless'') dust orbits at the full Keplerian speed.
As dust sediments into a thin sheet and creates a vertical shear, we expect that
Kelvin-Helmholtz instabilities might potentially develop when the local
dust-to-gas ratio $\mu\equiv\rho_d/\bar{\rho}_g$ in the midplane approaches
of order unity:
\begin{equation}\label{E:mu}
\mu_0\equiv\mu(z\!=\!0) \sim 1 \quad\mathrm{when}\quad \lambda \equiv H_d/H_g \sim \Sigma_d/\Sigma_g,
\end{equation}
where $\rho_d$ is the local mass density of dust, $H_d$ is the scale height
of the dust sub-layer, $\Sigma_d$ is the surface mass density of dust and
$\Sigma_g$ is the surface mass density of gas. When Kelvin-Helmholtz
instabilities do develop, the horizontal and vertical scales of the fastest
growing unstable modes are typically of order the thickness of the shear
layer:
$(\Delta r,r\Delta\phi,\Delta z)\sim H_d\sim\lambda H_g$.
We can divide the velocity into two components: the velocity across the
region of interest due to the Keplerian shear $\bar{v}\sim H_d\Omega_K\sim
\lambda H_g\Omega\sim\lambda c_s$, and the differential velocity between
the settled dust sub-layer and the dust-depleted gas above and below the
midplane $\tilde{v}\sim\eta V_K\sim\delta c_s$.  The Mach number of the
flow is thus: $\epsilon = \max(\delta,\lambda)$.

Motivated by this dimensional analysis, we simulate the dynamics only
within a small patch of the disk
$(r-r_0,r_0(\phi-\phi_0),z) \rightarrow (x,y,z) \sim H_d \sim \lambda H_g$
that co-rotates with the gas at some fiducial radius $r_0$ with angular speed
$\Omega_F\equiv\Omega_{K0}(1-\eta_0)$, where $\Omega_{K0}\equiv\Omega_K(r_0)$
and $\eta_0\equiv\eta(r_0,z\!=\!0)$.
The tidal term (\ie the remainder after the near cancellation of the
inward radial protostellar gravity and the outward centrifugal force)
and the equilibrium pressure gradient are given by:
\begin{mathletters}
\begin{eqnarray}\label{E:base_flow}
-\frac{\partial\Phi}{\partial r} + \Omega_F^2r & = & \Omega_{K0}^2r_0\left[3x/r_0-2\eta_0+\mathcal{O}\left(\delta^2\lambda^2,\delta^3\lambda,\delta^4\right)\right],\label{E:gravity_centrifugal_balance}\\
-\frac{1}{\bar{\rho}_g}\frac{\partial\bar{p}_g}{\partial r} & = & \Omega_{K0}^2r_0\left[2\eta_0 +\mathcal{O}\left(\delta^2\lambda^2,\delta^3\lambda\right)\right],\label{E:pressure_gradient_radial}\\
-\frac{1}{\bar{\rho}_g}\frac{\partial\bar{p}_g}{\partial z} & = & \Omega_{K0}^2r_0\left[z/r_0 +\mathcal{O}\left(\delta^2\lambda^2\right)\right].\label{E:pressure_gradient_vertical}
\end{eqnarray}
\end{mathletters}
\noindent In this work, we assume the background gas temperature is spatially constant, $\bar{T} = T_0$, so that the gas pressure gradient can be written in terms of the gas density gradient: $(\mathbf{\nabla}\bar{p}_g )/\bar{\rho}_g = \mathcal{R}T_0\mathbf{\nabla}\ln\bar{\rho}_g$, where $\mathcal{R}\equiv C_P-C_V$ is the gas constant.  The equilibrium gas density is thus:
\begin{mathletters}
\begin{eqnarray}
\bar{\rho}_g(x,z) &=& \rho_0\exp\left(-x/\Lambda_g-z^2/2H_g^2\right),\\
H_g^2 &\equiv& \mathcal{R}T_0/\Omega_{K0}^2,\\
\Lambda_g &\equiv &\mathcal{R}T_0/2\Omega_{K0}^2\eta_0r_0 = (c_{s0}/2\eta_0V_{K0})H_g.
\end{eqnarray}
\end{mathletters}

\subsection{Dynamic Equations for Gas \& Dust}

The Euler equations for perfectly coupled gas and dust in the 3D Cartesian shearing box are:
\begin{mathletters}\label{E:dynamic_equations1}
\begin{eqnarray}
\frac{d\mathbf{v}}{dt} & = & - 2\Omega_{K0}\mathbf{\hat{z}}\times\mathbf{v}  + \left(3\Omega_{K0}^2x-2\Omega_{K0}^2r_0\eta_0\right)\mathbf{\hat{x}} - \Omega_{K0}^2z\mathbf{\hat{z}} - \frac{1}{(\rho_g+\rho_d)}\mathbf{\nabla}p_g,\\
\frac{d\rho_g}{dt} & = & -\rho_g\mathbf{\nabla}\cdot\mathbf{v},\\
\frac{d\rho_d}{dt} & = & -\rho_d\mathbf{\nabla}\cdot\mathbf{v},\\
\frac{ds_g}{dt} & = & 0, \quad s_g \equiv C_V\ln(p_g\rho_g^{-\gamma}) + s_{g0}
\end{eqnarray}
\end{mathletters}
\noindent where $\mathbf{v}$ is the velocity of a parcel of gas and dust, $p_g$, $\rho_g$, and $s_g$ are the pressure, density, and entropy of the gas; $\rho_g$ is the density of the dust;  $\gamma\equiv C_P/C_V$ is the ratio of specific heats at constant pressure and constant volume; the advective or Lagrangian derivative is defined $d/dt\equiv(\partial/\partial t + \mathbf{v\cdot\nabla)}$.  The key difference between the dynamics of the gas and that of the dust is that we treat the dust as a cold, pressureless fluid.

The dust continuity equation can be recast in terms of the local dust-to-gas ratio: $d\mu/dt=0$, that is, the local dust-to-gas ratio is an advectively conserved quantity, meaning that a parcel of fluid maintains its dust content. This is simply a consequence of the perfect coupling assumption which does not allow the dust component to slip apart from the gas component.  This approximation is valid if the stopping time $t_S$ (the $e$-folding time for a particle's velocity to match that of the surrounding medium because of frictional coupling) is much shorter than the timescales of interest for Kelvin-Helmholtz instability.  For small particles for which the gas mean-free-path is larger than the size of the particles, Epstein drag sets this timescale \citep{cuzzi93,garaud04a}:
\begin{equation}\label{E:epstein}
t_{S,epstein}/t_{orb} \equiv \frac{\rho_s a/\rho_gc_s}{2\pi/\Omega_K} \sim 10^{-3}~\left(\frac{a}{1~\textrm{cm}}\right)\left(\frac{r}{1~\textrm{AU}}\right)^{3/2},
\end{equation}
where $\rho_s$ and $a$ are the solid density and radius of a dust grain, and $\rho_g$ and $c_s$ are the gas density and sound speed.

Because we expect the Kelvin-Helmholtz instability to set in at low Mach number $\epsilon\sim\max(\delta,\lambda)\ll 1$, we invoke the anelastic approximation for the gas flow.  We decompose the gas pressure and density into their equilibrium components (denoted with overbars) and fluctuating components (denoted with tildes): $p_g = \bar{p}_g + \tilde{p}_g$, $\rho_g = \bar{\rho}_g + \tilde{\rho}_g$.  At low Mach number, the fluctuating components should scale as: $\tilde{p}_g/\bar{p}_g \sim \tilde{\rho}_g/\bar{\rho}_g \sim \epsilon^2 <<1$.  The gas pressure gradient and gas continuity equations can then be expanded:
\begin{mathletters}\label{E:anelastic_approximation}
\begin{eqnarray}
\frac{1}{\rho_g}\mathbf{\nabla}p_g &\approx& \left[\frac{1}{\bar{\rho}_g}\mathbf{\nabla}\bar{p}_g +  \frac{1}{\bar{\rho}_g}\mathbf{\nabla}\tilde{p}_g - \frac{\tilde{\rho}_g}{\bar{\rho}_g}\frac{1}{\bar{\rho}_g}\mathbf{\nabla}\bar{p}_g\right]\left[1+\mathcal{O}(\epsilon^4)\right],\label{E:anelastic_pressure}\\
0 &\approx& \left[\mathbf{\nabla}\cdot\left(\bar{\rho}_g\mathbf{v}\right)\right]\left[1 + \mathcal{O}(\epsilon^2)\right].\label{E:anelastic_continuity}
\end{eqnarray}
\end{mathletters}
\noindent The anelastic approximation has been used extensively in the study of deep,
subsonic convection in planetary atmospheres \citep{ogura62,gough69} and
stars \citep{gilman81,glatzmaier81a,glatzmaier81b}. \citet{barranco00a} and \citet{barranco00b,barranco05,barranco06} previously used the anelastic approximation to study 3D vortices in protoplanetary disks.  One of the consequences of this approximation is that the total density is replaced by the time-independent mean density in the mass continuity equation, which has the effect of filtering high-frequency acoustic waves and shocks, but allowing slower wave phenomena such as internal gravity waves.


The dynamic equations for coupled gas and dust with the constant temperature background and anelastic approximation become:
\begin{mathletters}\label{E:dynamic_equations}
\begin{eqnarray}
\frac{d\mathbf{v}}{dt} & = & - 2\Omega_{K0}\mathbf{\hat{z}}\times\mathbf{v}  + 3\Omega_{K0}^2x\mathbf{\hat{x}} + \frac{\tilde{T}}{T_0}\left(2\Omega_{K0}^2r_0\eta_0\mathbf{\hat{x}}+\Omega_{K0}^2z\mathbf{\hat{z}}\right) - \mathbf{\nabla}\tilde{h}_g \nonumber\\
{} & {} & - \frac{\mu}{\mu+1}\left[\left(1+\frac{\tilde{T}}{T_0}\right)\left(2\Omega_{K0}^2r_0\eta_0\mathbf{\hat{x}}+\Omega_{K0}^2z\mathbf{\hat{z}}\right)-\mathbf{\nabla}\tilde{h}_g\right],\\
0 & = & \mathbf{\nabla}\cdot\bar{\rho}_g\mathbf{v},\\
\frac{d\mu}{dt} & = & 0, \quad \mu \equiv \rho_d/\bar{\rho_g},\\
C_P\frac{d\tilde{T}}{dt} & = & -\left(1+\frac{\tilde{T}}{T_0}\right)\mathbf{v}\cdot\left(2\Omega_{K0}^2r_0\eta_0\mathbf{\hat{x}}+\Omega_{K0}^2z\mathbf{\hat{z}}\right),\\
\tilde{p}_g &=& \tilde{\rho}_g\mathcal{R}T_0 + \bar{\rho}_g\mathcal{R}\tilde{T}, \quad \tilde{h}_g \equiv \tilde{p}_g/\bar{\rho}_g.
\end{eqnarray}
\end{mathletters}
\noindent When the background temperature is spatially constant,  the gas enthalpy turns out to be be a more useful quantity than the gas pressure: $\tilde{h}_g\equiv\tilde{p}_g/\bar{\rho}_g$. These equations are almost identical to the anelastic equations in \citet{barranco05,barranco06}, with the addition of a nonlinear forcing term for the inertia of the dust.

The above set of dynamic equations allow the following steady-state equilibrium (denoted with the dagger symbol):
\begin{mathletters}\label{E:equilibrium}
\begin{eqnarray}
0 &=& v_x^{\dagger} = v_z^{\dagger} = \tilde{T}^{\dagger},\\
\mu^{\dagger}(z) &{}& \; \mathrm{arbitrary},\\
v_y^{\dagger}(x,z) &=& -\case{3}{2}\Omega_{K0}x + \left[\frac{\mu^{\dagger}(z)}{\mu^{\dagger}(z)+1}\right]\Omega_{K0}r_0\eta_0,\\
\frac{\partial\tilde{h}_g^{\dagger}(z)}{\partial z} & = & -\mu^{\dagger}(z)\Omega_{K0}^2z.
\end{eqnarray}
\end{mathletters}

In the limit where we take $\Lambda_g\! \rightarrow\! \infty$ and neglect the radial variation of the background gas density and radial component of the gas buoyancy, one can derive the following global energy balance equations:
\begin{mathletters}\label{E:energy}
\begin{eqnarray}
KE &\equiv & \int_V \case{1}{2}(1+\mu)\bar{\rho}_g\mathbf{v\cdot v}\; dV,\label{E:KE}\\
PE &\equiv & \int_V \left[\bar{\rho}_g\mu\Omega_{K0}^2\left(2r_0\eta_0x + \case{1}{2}z^2 - \case{3}{2}x^2\right) + C_P\bar{\rho}_g\tilde{T}\right]\; dV\label{E:PE},\\
\frac{d}{dt}\left(KE+PE\right) & = & \case{3}{2}\Omega_{K0}L_x\int_S\left[(1+\mu)\bar{\rho}_g\tilde{v}_x\tilde{v}_y\right]\vert_{x=+L_x/2}\; dydz  \nonumber \\ {} & {} & -2\Omega_{K0}^2r_0\eta_0L_x\int_S\left[\bar{\rho}_g\mu v_x\right]\vert_{x=+L_x/2}\; dydz,
\end{eqnarray}
\end{mathletters}

\subsection{Brief description of numerical method}

Here we briefly describe the numerical method; a more detailed presentation can be found in \citet{barranco06}.  We solve the dynamic equations~\eqref{E:dynamic_equations} with a spectral method; that is, each variable is represented as a finite sum of basis functions multiplied by spectral coefficients \citep{gottlieborszag77,marcus86a,canuto88,boyd89}.  The choice of basis functions for each direction is guided by the corresponding boundary conditions.  The equations are autonomous in the azimuthal coordinate $y$, so it is reasonable to assume periodic boundary conditions in this direction.  However, the equations explicitly depend on the radial coordinate $x$ because of the linear background shear.  We adopt ``shearing box'' boundary conditions:  $q(x+L_x,y-(3/2)\Omega_0L_xt,z,t) = q(x,y,z,t)$, where $q$ represents any of $\mathbf{v}$, $\tilde{h}$, $\tilde{T}$, etc.   In practice, we rewrite the equations~\eqref{E:dynamic_equations} in terms of quasi-Lagrangian or shearing coordinates that advect with the background shear \citep{goldreich65b,marcus77,rogallo81}:  $t' \equiv  t$, $x' \equiv  x$,  $y' \equiv  y + (3/2)\Omega_{0} x t$, and $z' \equiv  z$.  In these new coordinates, the radial boundary conditions become: $q(x'+L_x,y',z',t') = q(x',y',z',t')$.  That is, shearing box boundary conditions are equivalent to periodic boundary conditions in the  shearing coordinates.  Physically, this means that the periodic images at different radii are not fixed, but advect with the background shear.

In the shearing coordinates, the equations are autonomous in both $x'$ and $y'$ (although they now explicitly depend on $t'$), making a Fourier basis the natural choice for the spectral expansions in the horizontal directions:   
\begin{equation}\label{E:spectral}
q(x',y',z',t') = \sum_{\mathbf{k}} \hat{q}_{\mathbf{k}}(t')e^{ik'_xx'}e^{ik'_yy'}\phi_n(z'),
\end{equation}
where $q$ is any variable of interest, $\{\hat{q}_{\mathbf{k}}(t')\}$ is the set of spectral coefficients, and $\mathbf{k}=\{k'_x,k'_y,n\}$ is the set of wavenumbers.  We have implemented the simulations with two different sets of basis functions for the spectral expansions in the vertical direction, corresponding to two different sets of boundary conditions: 

\noindent (i) For the truncated domain $-L_z\le z \le L_z$, we use Chebyshev polynomials: $\phi_n(z) = T_n(z/L_z) \equiv \cos(n\xi)$, where $\xi\equiv\cos^{-1}(z/L_z)$.  We apply the condition that the vertical velocity vanish at the top and bottom boundaries: $v_z(x,y,z\!=\!\pm L_z,t) = 0$.

\noindent (ii) For the infinite domain  $-\infty < z < \infty$, we use rational Chebyshev functions: $\phi_n(z) = \cos(n\xi)$ (for $v_x$, $v_y$, and all the thermodynamic variables) or $\phi_n(z)=\sin(n\xi)$ (for $v_z$), where $\xi \equiv \cot^{-1}(z/L_z)$.  In this context, $L_z$ is no longer the physical size of the box, but is a mapping parameter; exactly one half of the grid points are within $|z|\le L_z$, whereas the other half are widely spaced in the region $L_z < |z| < \infty$.    No explicit boundary conditions on the vertical velocity are necessary when we solve the equations on the infinite domain because the $\phi_n(z)=\sin(n\xi)$ basis functions individually decay to zero at large $z$.

The equations are integrated forward in time with a fractional step method:  the nonlinear advection terms are integrated with an explicit second-order Adams-Bashforth method, and the pressure step is computed with a semi-implicit second-order Crank-Nicholson method.  The time integration scheme is overall globally second-order accurate.  Unlike finite-difference methods, spectral methods have no inherent grid dissipation; energy cascades to smaller and smaller size scales via the nonlinear interactions, where it can ``pile-up'' and potentially degrade the convergence of the spectral expansions.  We employ a $\nabla^{12}$ hyperviscosity or low-pass filter every timestep to damp the energy at the highest wavenumbers.

Different horizontal Fourier modes interact only through the nonlinear advective terms; once these terms are computed, the horizontal Fourier modes can be decoupled.  This motivated us to parallelize the code: each processor computes on a different block of data in horizontal Fourier wavenumber space.  Parallelization is implemented with Message Passing Interface (MPI),  typically using between 64 and 512 processors.  Wall-clock time scales inversely with number of processors, indicating near-optimal parallelism; timing analyses are presented in \citet{barranco06}.   

\section{REVISITING THE CASE OF NO DIFFERENTIAL ROTATION}

The linear stability of settled dust layers to Kelvin-Helmholtz instability has previously been investigated by a number of researchers \citep{sekiya98,sekiya00,youdin02,garaud04,gomez05}; however, almost all of these analyses neglected the role of the differential rotation in the protoplanetary disk.  If one were to directly linearize the equations \eqref{E:dynamic_equations}, one would find the resulting set to depend linearly on the radial coordinate $x$ because of the shear, making it difficult to apply periodic boundary conditions in the radial direction.  Alternatively, one could employ a set of coordinates that advect with the background shear \citep{goldreich65b,marcus77,ryu92}, but then the resulting linearized equations would explicitly depend on time.  \citet{ishitsu03} employed this approach using an initial-value code to address the effect of horizontal shear on unstable eigenmodes in the small-amplitude, linear regime.  They found that modes grew for a limited period of time, but were eventually stabilized as they were sheared out to high spatial wavenumber.  No matter the approach, the linear stability analysis is difficult to treat analytically, motivating theorists to tackle the problem without the background shear with the hope that the results are qualitatively, if not quantitatively, useful in determining whether or not the Kelvin-Helmholtz instability is a barrier to further settling.

In this section, we briefly revisit the case of no radial shear, first without, and then with the Coriolis force.  
We will neglect the $x$ (radial) dependence of the background gas density (setting $\Lambda_g \!\rightarrow\! \infty$), which eliminates the radial component of gas buoyancy.  We assume eigenmodes of the form $q'(t,x,y,z)=q'(z)\exp(-i\omega t+ik_xx+ik_yy)$.  The linearized equations corresponding to \eqref{E:dynamic_equations}, \textit{without} radial shear and radial gas buoyancy, are:
\begin{mathletters}\label{E:linearized}
\begin{eqnarray}
-i\omega v'_x &=& -v_y^{\dag}(z)ik_yv'_x + 2\Omega_{K0}v'_y - \left[\frac{ik_x}{[1+\mu^{\dag}(z)]}\right]h' - \left[\frac{2\Omega^2_{K0}\eta_0r_0}{[1+\mu^{\dag}(z)]^2}\right]\mu',\\
-i\omega v'_y &=& -2\Omega_{K0}v'_x -v_y^{\dag}(z)ik_yv'_y - \left[\frac{dv_y^{\dag}}{dz}\right]v'_z - \left[\frac{ik_y}{[1+\mu^{\dag}(z)]}\right]h',\\
-i\omega v'_z &=& -v_y^{\dag}(z)ik_yv'_z - \left[\frac{(\partial/\partial z)}{[1+\mu^{\dag}(z)]}\right]h' + \left[\frac{\Omega_{K0}^2z}{[1+\mu^{\dag}(z)]}\right]\frac{T'}{T_0} - \left[\frac{\Omega_{K0}^2z}{[1+\mu^{\dag}(z)]}\right]\mu',\\
-i\omega\frac{T'}{T_0} &=& -v_y^{\dag}(z)ik_y\frac{T'}{T_0} + \left[\frac{\mathcal{R}}{C_P}\frac{d\ln\bar{\rho}_g}{dz}\right]v'_z,\\
-i\omega \mu' &=& -v_y^{\dag}(z)ik_y\mu' - \left[\frac{d\mu^{\dag}}{dz}\right]v'_z,\\
0 &=& ik_xv'_x+ik_yv'_y+\left[\frac{\partial}{\partial z} + \frac{d\ln\bar{\rho}_g}{dz}\right]v'_z.
\end{eqnarray}
\end{mathletters}
\noindent One can eliminate $T'$ and $\mu'$ from the equation for $v'_z$, yielding:
 \begin{equation}
 \left[-i\omega + v_y^{\dag}(z)ik_y\right]^2v'_z = -\left[\frac{-i\omega + v_y^{\dag}(z)ik_y}{[1+\mu^{\dag}(z)]}\right]\frac{\partial h'}{\partial z} + \frac{\Omega_{K0}^2z}{[1+\mu^{\dag}(z)]}\left(\frac{\mathcal{R}}{C_P}\frac{d\ln\bar{\rho}_g}{dz} + \frac{d\mu^{\dag}}{dz}\right)v'_z.
 \end{equation}
 This form allows us to easily identify the Brunt-V\"{a}is\"{a}l\"{a} frequency:
 \begin{equation}\label{E:brunt}
 N^2 = -\frac{\Omega_{K0}^2z}{[1+\mu^{\dag}(z)]}\left(\frac{\mathcal{R}}{C_P}\frac{d\ln\bar{\rho}_g}{dz} + \frac{d\mu^{\dag}}{dz}\right).
 \end{equation}
 
 \begin{figure}
\epsscale{1.0}
\plotone{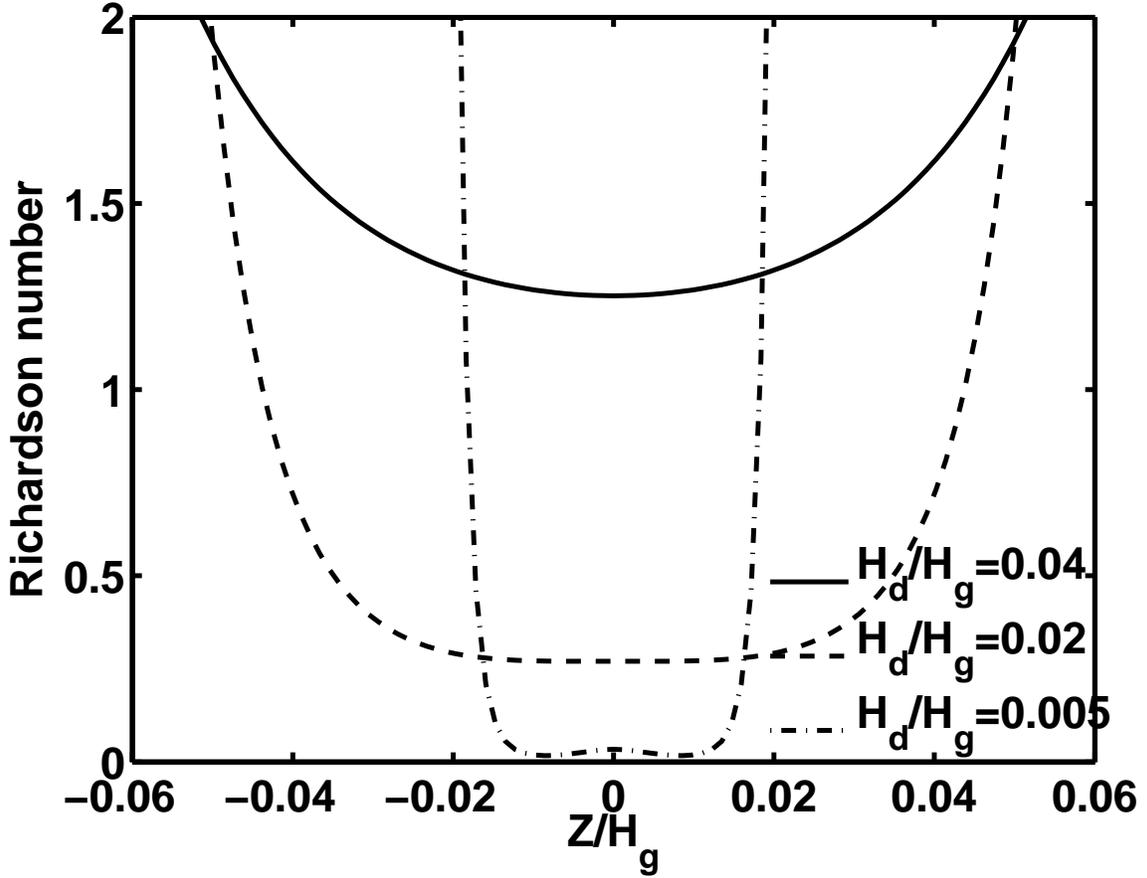}
\caption{\label{F:Ri_gauss} Gradient Richardson number \eqref{E:gradient_richardson} for Gaussian profiles of the local dust-to-gas ratio $\mu$.  For these three curves, $\Sigma_d/\Sigma_g = 0.01$, $\eta_0V_{K0}/c_{s0} = 0.1$, and $H_d/H_g = 0.04, 0.02, 0.005$, corresponding to peak dust-to-gas-ratios of $\mu^{\dag}_0= 0.25, 0.50, 2.0$.  For $\mu^{\dag}_0=1/2$, the Richardson number is at a minimum right at the midplane ($z\!=\!0$) and is nearly constant throughout much of the dust layer before rising sharply at the edge of the dust distribution.  For $\mu^{\dag}_0>1/2$, the Richardson number is still relatively constant throughout the core of the layer, although the minimum has shifted off the midplane.   }
\end{figure}

Before proceeding with the linear analysis, we must specify the form of the vertical distribution of dust.  We choose a Gaussian profile for the local dust-to-gas ratio:
 \begin{equation}\label{E:gaussian_dust}
 \mu^{\dag}(z) = \mu^{\dag}_0\exp\left(-z^2/2H_{\mu}^2\right), \quad \mu^{\dag}_0\equiv\frac{\Sigma_d}{\Sigma_g}\frac{H_g}{H_d},\quad H_{\mu}^{-2} \equiv H_d^{-2} - H_g^{-2},
 \end{equation}
where we have defined the midplane dust-to-gas ratio $\mu^{\dag}_0$,  the initial Gaussian scale height for the dust density $H_d$, and the initial Gaussian scale height for the dust-to-gas ratio $H_{\mu}$.
The gradient Richardson number \eqref{E:Richardson_criterion} for a Gaussian distribution of dust is:
\begin{equation}\label{E:gradient_richardson}
Ri(z) = \left(\frac{\eta_0 V_{K0}}{c_{s0}}\right)^{-2}\left(\frac{H_{\mu}}{H_g}\right)^2\frac{[1+\mu^{\dag}(z)]^3}{\mu^{\dag}(z)}\left[1 + \frac{\mathcal{R}}{C_P}\frac{1}{\mu^{\dag}(z)}\left(\frac{H_{\mu}}{H_g}\right)^2\right].
\end{equation}
In Figure~\ref{F:Ri_gauss}, we have graphed the gradient Richardson number with $\Sigma_d/\Sigma_g = 0.01$ and $\eta_0V_{K0}/c_{s0} = 0.1$, for a few different dust scale heights.  For $\mu^{\dag}_0=1/2$ ($H_d/H_g = 0.02$), the Richardson number is at a minimum right at the midplane ($z\!=\!0$) and is nearly constant throughout much of the dust layer before rising sharply at the edge of the dust distribution.  For $\mu^{\dag}_0>1/2$ ($H_d/H_g < 0.02$), the Richardson number is still relatively constant throughout the core of the layer, although the minimum has shifted off the midplane.  For very thin dust layers, we can ignore the gas buoyancy near the midplane (\ie the $\mathcal{R}/C_P$ term on the far right of equation~\eqref{E:gradient_richardson}); the minimum gradient Richardson number is thus:
\begin{equation}\label{E:minimum_richardson}
Ri_{min} =  \left(\frac{\eta_0 V_{K0}}{c_{s0}}\right)^{-2}\left(\frac{H_{\mu}}{H_g}\right)^2 \times
\left\{ \begin{array}{l}
(27/4) \quad\mbox{at}\quad z = \pm H_{\mu}\sqrt{2\ln(2\mu^{\dag}_0)}\quad\mbox{for}\quad\mu^{\dag}_0>1/2\\ (1+\mu^{\dag}_0)^3/\mu^{\dag}_0\quad\mbox{at}\quad z=0 \quad\mbox{for}\quad\mu^{\dag}_0<1/2.
\end{array}\right.
\end{equation}
It is interesting to note that for dust-rich layers ($\mu^{\dag}_0>1/2$), the minimum Richardson number is independent of dust-to-gas ratio.

We numerically solve the eigenproblem \eqref{E:linearized} for the complex frequencies $\omega$ with a Chebyshev spectral method in which the top and bottom boundaries are mapped to infinity \citep{barranco06,boyd89,cain84}.   Because we are neglecting differential rotation, we invoke Squire's theorem which states that two-dimensional eigenmodes are more unstable than three-dimensional ones \citep{squire33,chandra61,drazin81}, and so set $k_x=0$.  We also restrict our analysis to modes for which the dust-to-gas ratio perturbation $\mu'$ is an odd-function of the vertical coordinate $z$.  

\subsection{Case of no Coriolis force and no horizontal shear}

\begin{figure}
\epsscale{1.0}
\plotone{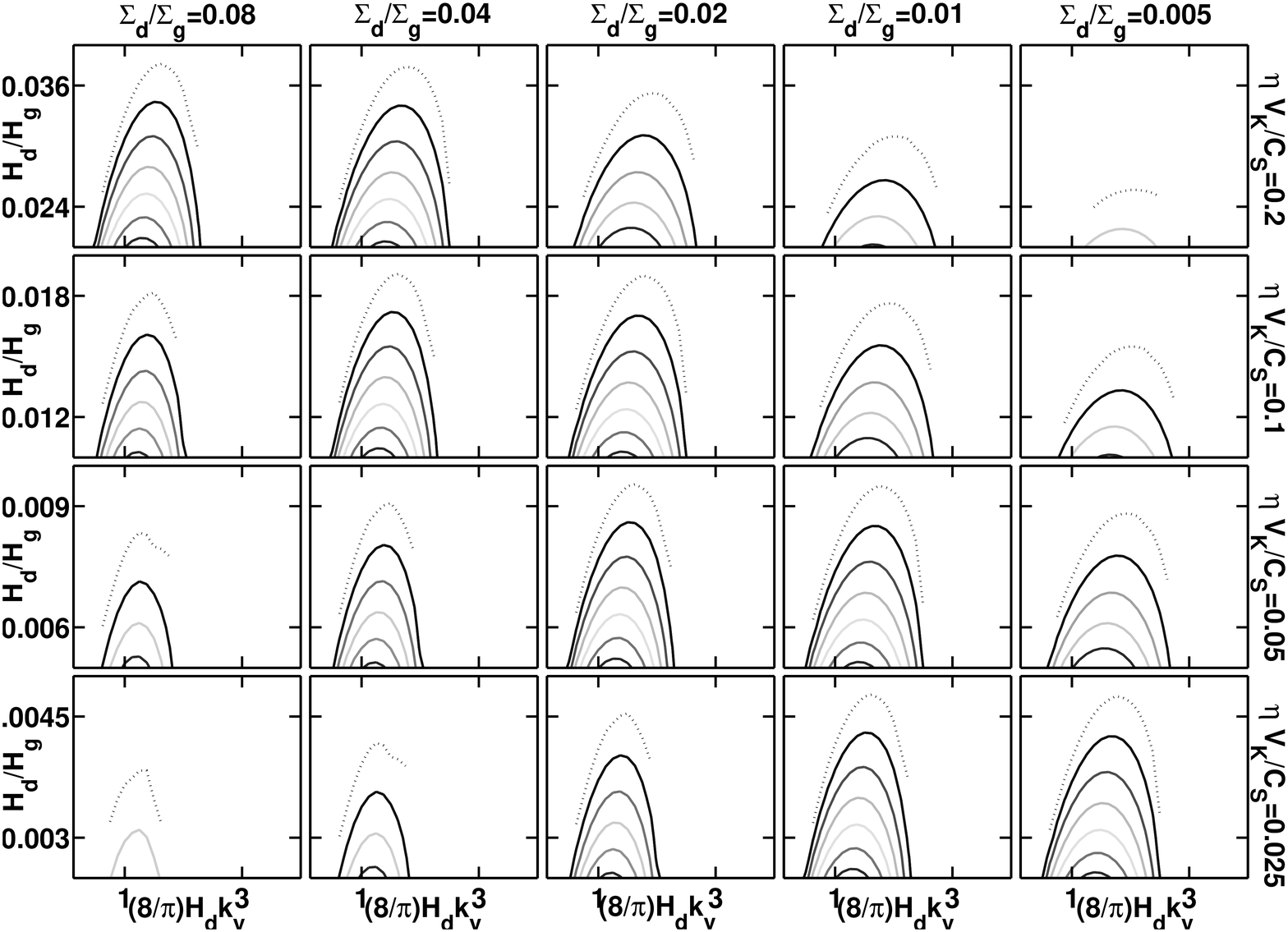}
\caption{\label{F:growth_nocoriolis} Contour plots of growth rates of Kelvin-Helmholtz instability for the case of no Coriolis force and no horizontal shear.  Global dust-to-gas ratio varies across rows with values $\Sigma_d/\Sigma_g$ = 0.08, 0.04, 0.02, 0.01, 0.005.  Strength of radial pressure gradient varies down columns with values $\eta_0V_{K0}/c_{s0}$ = 0.2, 0.1, 0.05, 0.025.  The horizontal axis of each plot is the nondimensionalized wavenumber $(8/\pi)H_d k_y$, and the vertical axis is the ratio of the dust scale height to the gas scale height.  The solid contours, from outer to inner, correspond to growth rates of  0.1, 0.2, 0.3, 0.4, 0.5, 0.6 in units of $\Omega_{K0}^{-1}$. The outermost, dotted contour corresponds to an extrapolation to zero growth rate.}
\end{figure}

\begin{figure}
\epsscale{1.0}
\plotone{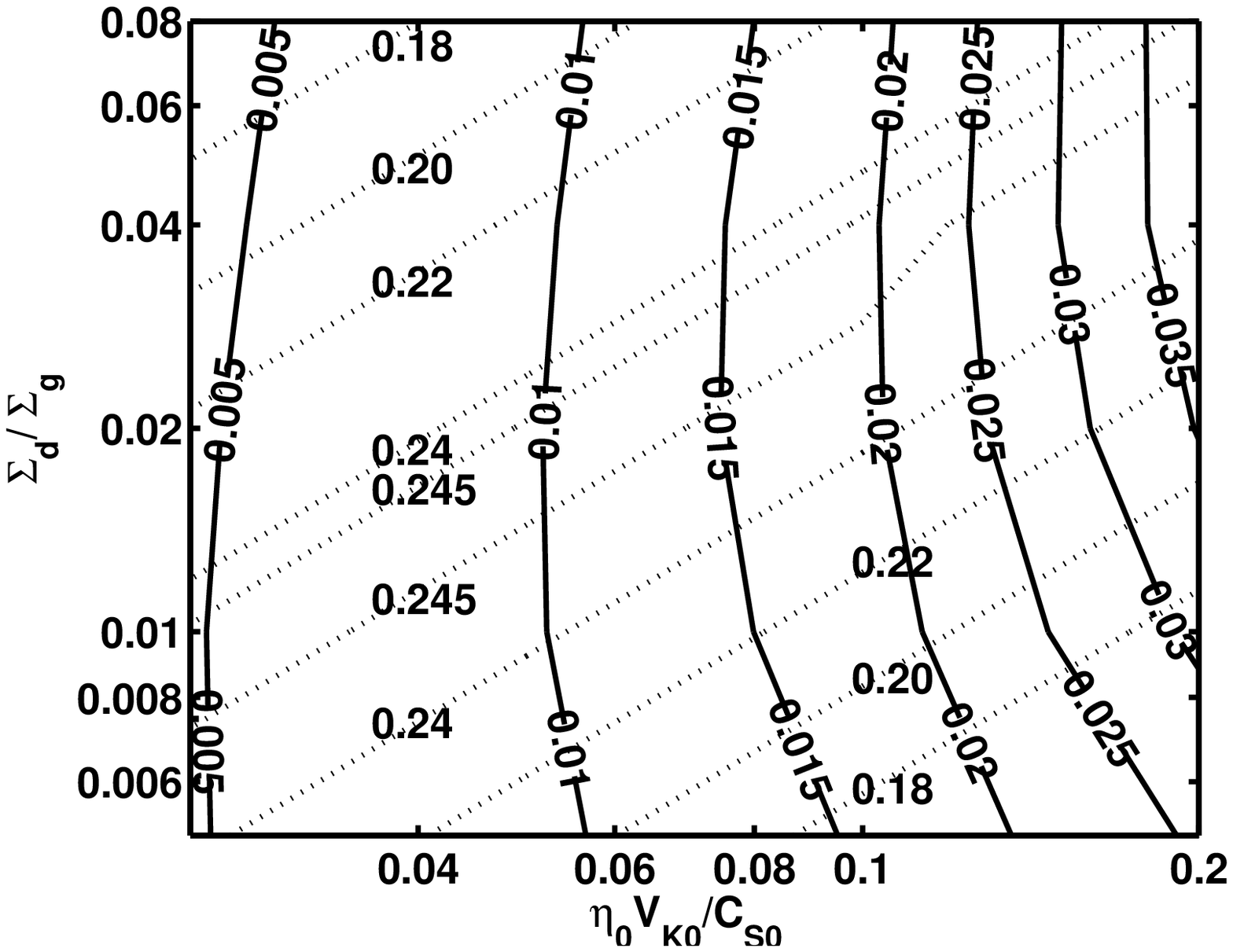}
\caption{\label{F:Ri_crit} Contours in the $(\Sigma_d/\Sigma_g,\eta_0V_{K0}/c_{s0})$ plane for the minimum dust layer thickness $H_d/H_g$ (solid black lines) and minimum Richardson number $Ri_{min}$ (dotted lines) at the onset of instability.  Over the range of parameter space explored, the minimum Richardson number is in the range 0.18 to 0.25, as expected for ``classic'' Kelvin-Helmholtz instability with no Coriolis force and no horizontal shear.  Note how the contours for $H_d/H_g$ are nearly vertical, indicating a weak dependence on the global dust-to-gas ratio for the onset of instability.  This figure is consistent with Figure~11 in \citet{garaud04}.}
\end{figure}

Figure~\ref{F:growth_nocoriolis} shows the growth rates (imaginary part of the complex eigenvalue $\omega$) for the case where the Coriolis force is turned-off.  Each of the twenty plots corresponds to different values of the global dust-to-gas ratio, $\Sigma_d/\Sigma_g$, and the strength of the radial gas pressure gradient, $\eta_0V_{K0}/c_{s0}$, which sets the maximum differential velocity between pure dust and pure gas .  Global dust-to-gas ratio varies across rows with values $\Sigma_d/\Sigma_g$ = 0.08, 0.04, 0.02, 0.01, 0.005.  Strength of radial pressure gradient varies down columns with values $\eta_0V_{K0}/c_{s0}$ = 0.2, 0.1, 0.05, 0.025.  The horizontal axis of each individual plot is the nondimensionalized wavenumber $(8/\pi)H_d k_y$, and the vertical axis is the ratio of the dust scale height to the gas scale height.  The solid contours, from outer to inner, correspond to growth rates of  0.1, 0.2, 0.3, 0.4, 0.5, 0.6 in units of $\Omega_{K0}^{-1}$.  The outermost (dotted) contour corresponds to an extrapolation to zero growth rate.

The peak of the zero-growth contour reveals the thinnest layer to remain stable to Kelvin-Helmholtz instability as well as the wavelength of the eigenmode at the onset of instability.  The wavelength at onset  is typically between 8 and 16 times the dust scale height over the range of parameter space explored.   In Figure~\ref{F:Ri_crit}, we plot contours in the $(\Sigma_d/\Sigma_g,\eta_0V_{K0}/c_{s0})$ plane for the minimum dust scale height $H_d/H_g$ (solid black lines) to remain stable.  We also plot contours (dotted lines) for the minimum gradient Richardson number~\eqref{E:minimum_richardson} corresponding to the minimum dust thickness at the onset of instability. Over the range of parameter space explored, the minimum Richardson number is in the range 0.18 to 0.25, as expected for ``classic'' Kelvin-Helmholtz instability with no Coriolis force and no horizontal shear.  Similar results were obtained by \citet{garaud04} (Figure~11 in their work).

\begin{figure}
\myplottwo{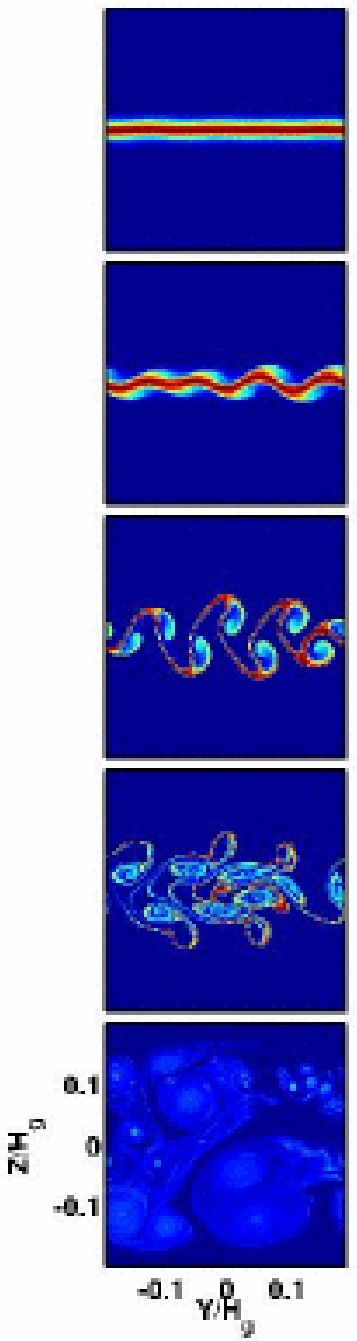}{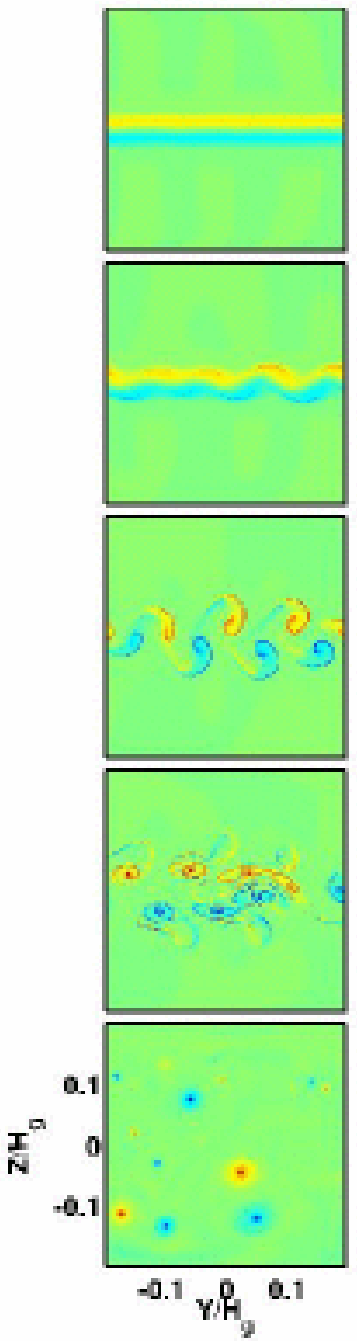}
\caption{\label{F:run28} Two-dimensional nonlinear evolution of Kelvin-Helmholtz instability with no Coriolis force and no horizontal shear. In this simulation, $\Sigma_d/\Sigma_g = 0.01$ and $\eta_0V_{K0}/c_{s0} = 0.1$.  The initial dust scale height is $H_d/H_g = 0.01$, corresponding to a peak local dust-to-gas ratio in the midplane of $\mu^{\dagger}_0 = 1$ and a minimum Richardson number of $Ri_{min} = 0.0675$.  The first column illustrates the evolution of the local dust-to-gas ratio $\mu$ (deep red = 1, deep blue = 0); the second column shows the evolution of the radial component of vorticity $\omega_x$ (red = vorticity that points into the page, blue = vorticity that points out of the page).  The times corresponding to each frame, in units of the orbital period, are: 3.8, 4.5, 5.1, 5.7, 15.3.  }
\end{figure}

We simulate the nonlinear evolution of the instability in order to investigate the nature of the subsequent mixing of gas and dust.  An example of a two-dimensional simulation in the $y-z$ plane with $\Sigma_d/\Sigma_g = 0.01$ and $\eta_0V_{K0}/c_{s0} = 0.1$ is presented in Figure~\ref{F:run28}.  The initial dust scale height is $H_d/H_g = 0.01$, corresponding to a peak local dust-to-gas ratio in the midplane of $\mu^{\dagger}_0 = 1$ and a minimum Richardson number of $Ri_{min} = 0.0675$.  The first column illustrates the evolution of the local dust-to-gas ratio $\mu$ (deep red = 1, deep blue = 0); the second column shows the evolution of the radial component of vorticity $\omega_x\equiv\partial v_z/\partial y - \partial v_y/\partial z$ (red = vorticity that points into the page, blue = vorticity that points out of the page).  The times corresponding to each frame, in units of the orbital period, are: 3.8, 4.5, 5.1, 5.7, 15.3.  The dust layer develops waves, which grow and break into pairs of anti-aligned vortices.   Characteristic of two-dimensional turbulence, like-signed vortices merge to form larger vortices as energy cascades to larger spatial scales.  These vortices chaotically interact, leading to thorough mixing of the dust with the gas. 

\subsection{Case with Coriolis force, but no horizontal shear}

\begin{figure}
\epsscale{1.0}
\plotone{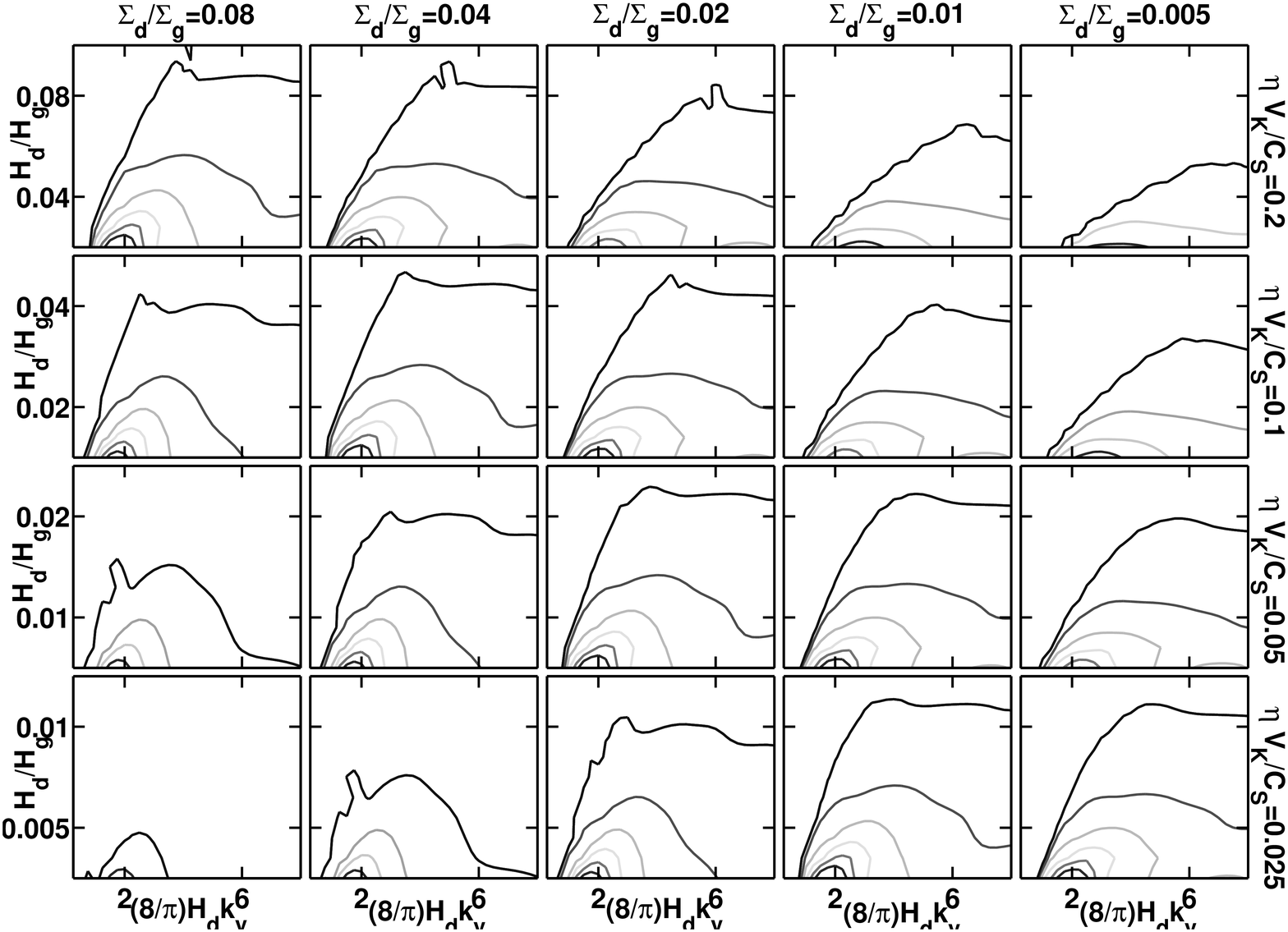}
\caption{\label{F:growth_coriolis} Contour plots of growth rates of Kelvin-Helmholtz instability for the case with the Coriolis force, but no horizontal shear.  Global dust-to-gas ratio varies across rows with values $\Sigma_d/\Sigma_g$ = 0.08, 0.04, 0.02, 0.01, 0.005.  Strength of radial pressure gradient varies down columns with values $\eta_0V_{K0}/c_{s0}$ = 0.2, 0.1, 0.05, 0.025.  The horizontal axis of each plot is the nondimensionalized wavenumber $(8/\pi)H_d k_y$, and the vertical axis is the ratio of the dust scale height to the gas scale height.  The solid contours, from outer to inner, correspond to growth rates of  0.1, 0.2, 0.3, 0.4, 0.5, 0.6 in units of $\Omega_{K0}^{-1}$.  Note that the horizontal and vertical axes of each plot have roughly twice the range as the corresponding plots in Figure~\ref{F:growth_nocoriolis} for the case of no Coriolis force.}
\end{figure}

Figure~\ref{F:growth_coriolis} shows the growth rates for the case with the Coriolis force, but no horizontal shear.  As before, each of the twenty plots corresponds to different values of the global dust-to-gas ratio, $\Sigma_d/\Sigma_g$, and the strength of the radial gas pressure gradient, $\eta_0V_{K0}/c_{s0}$, which sets the maximum differential velocity between pure dust and pure gas.  Global dust-to-gas ratio varies across rows with values $\Sigma_d/\Sigma_g$ = 0.08, 0.04, 0.02, 0.01, 0.005.  Strength of radial pressure gradient varies down columns with values $\eta_0V_{K0}/c_{s0}$ = 0.2, 0.1, 0.05, 0.025.  The horizontal axis of each individual plot is the nondimensionalized wavenumber $(8/\pi)H_d k_y$, and the vertical axis is the ratio of the dust scale height to the gas scale height.  The solid contours, from outer to inner, correspond to growth rates of  0.1, 0.2, 0.3, 0.4, 0.5, 0.6 in units of $\Omega_{K0}^{-1}$.  Note that the vertical scale of each plot is a factor of 2.2 larger than the corresponding ones in Figure~\ref{F:growth_nocoriolis}, and the horizontal scale is a factor of 2 larger.

\citet{gomez05} were the first to note that settled dust layers are more unstable when the Coriolis force is included.  Instability occurs for thicker layers and for a much larger range of wavenumbers.  In the case where there is no Coriolis force, the range of unstable wavenumbers for a given ratio of $H_d/H_g$ is relatively narrow.  In contrast, when the Coriolis force is included, we find that instability occurs for very large wavenumbers, with no apparent upper limit.  However, these large wavenumber (small wavelength) eigenmodes are the ones which will be most affected by the inclusion of horizontal shear.

\begin{figure}
\myplottwo{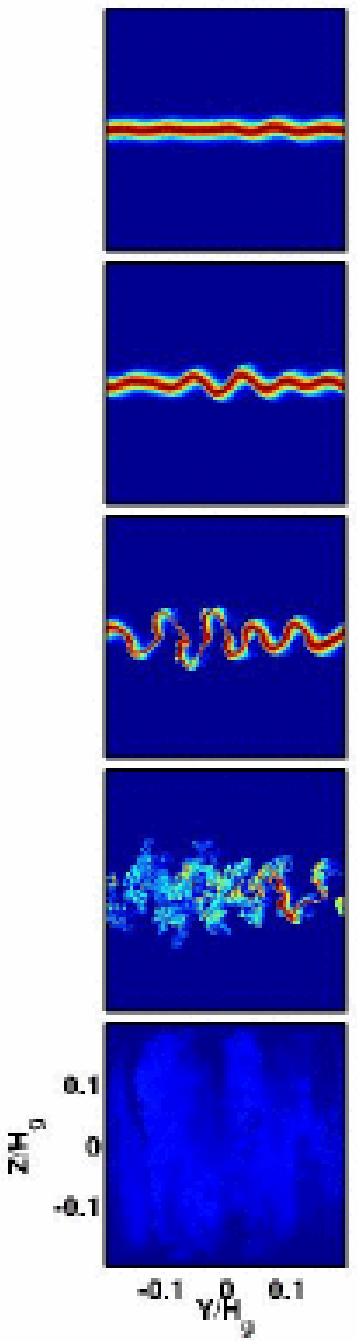}{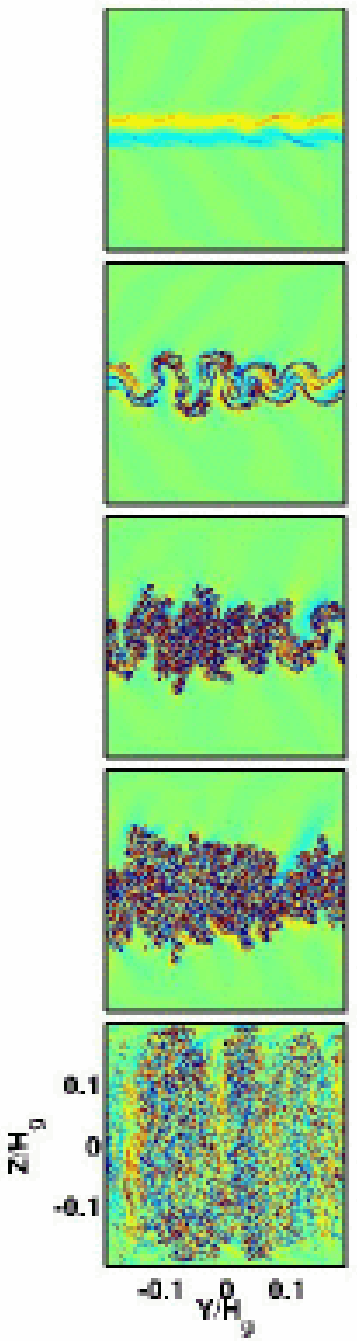}
\caption{\label{F:run29} Two-dimensional nonlinear evolution of Kelvin-Helmholtz instability with Coriolis force but still no horizontal shear. In this simulation, $\Sigma_d/\Sigma_g = 0.01$ and $\eta_0V_{K0}/c_{s0} = 0.1$.  The initial dust scale height is $H_d/H_g = 0.01$, corresponding to a peak local dust-to-gas ratio in the midplane of $\mu^{\dagger}_0 = 1$ and a minimum Richardson number of $Ri_{min} = 0.0675$.  The first column illustrates the evolution of the local dust-to-gas ratio $\mu$ (deep red = 1, deep blue = 0); the second column shows the evolution of the radial component of vorticity $\omega_x$ (red = vorticity that points into the page, blue = vorticity that points out of the page).  The times corresponding to each frame, in units of the orbital period, are: 1.9, 2.2, 2.5, 2.9, 7.6.  }
\end{figure}

Figure~\ref{F:run29} shows the two-dimensional nonlinear evolution of Kelvin-Helmholtz instability with Coriolis force but still no horizontal shear. This simulation is exactly the same as the one in Figure~\ref{F:run28}, except that the Coriolis force is included.  As before $\Sigma_d/\Sigma_g = 0.01$, $\eta_0V_{K0}/c_{s0} = 0.1$, $H_d/H_g = 0.01$, $\mu^{\dagger}_0 = 1$ and $Ri_{min} = 0.0675$.  The first column illustrates the evolution of the local dust-to-gas ratio $\mu$ (deep red = 1, deep blue = 0); the second column shows the evolution of the radial component of vorticity $\omega_x$ (red = vorticity that points into the page, blue = vorticity that points out of the page).  The times corresponding to each frame, in units of the orbital period, are: 1.9, 2.2, 2.5, 2.9, 7.6.  Waves appear on the dust layer as with the case with no Coriolis force, but no large-scale vortices develop.  The vorticity has more power at the smallest spatial scales.  The nonlinear mixing, however, is still very efficient, and the dust is completely re-mixed with the gas throughout the entire computational domain.

\begin{figure}
\myplottwo{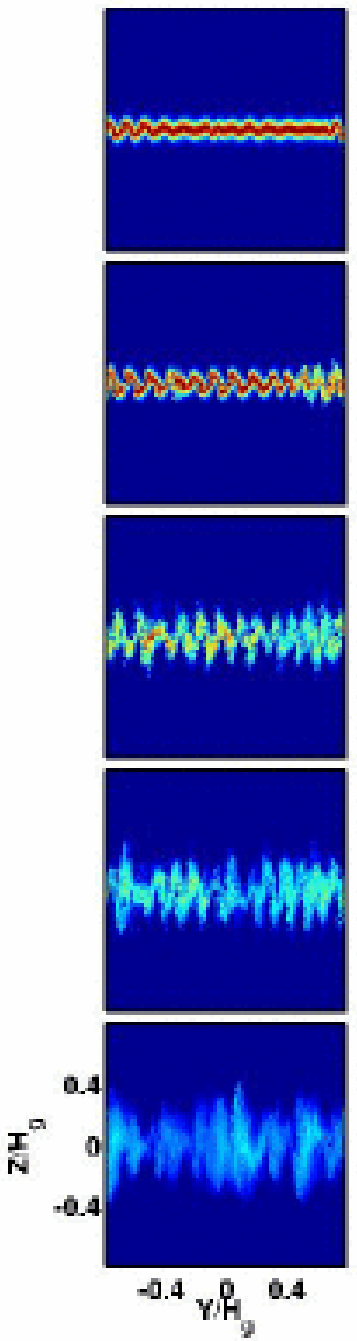}{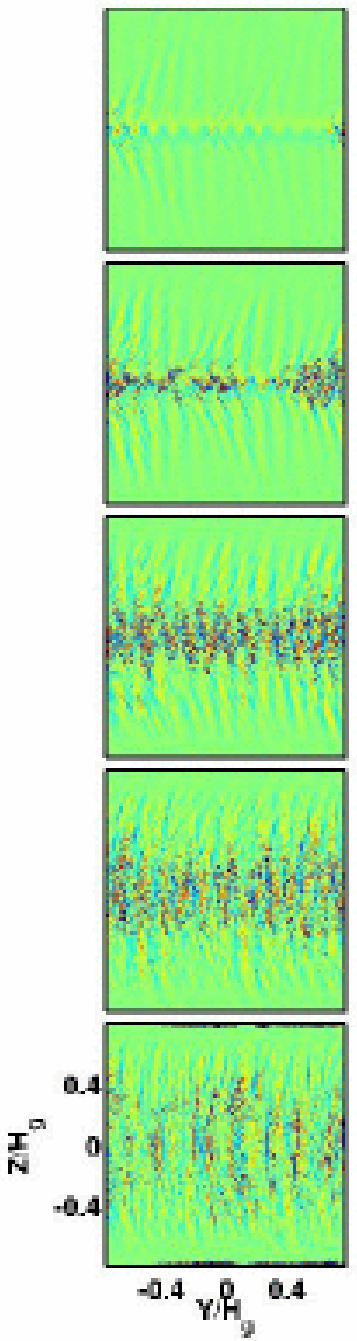}
\caption{\label{F:run30} Two-dimensional nonlinear evolution of Kelvin-Helmholtz instability with Coriolis force but still no horizontal shear. In this simulation, $\Sigma_d/\Sigma_g = 0.01$ and $\eta_0V_{K0}/c_{s0} = 0.1$.  The initial dust scale height is $H_d/H_g = 0.04$, corresponding to a peak local dust-to-gas ratio in the midplane of $\mu^{\dagger}_0 = 0.25$ and a minimum Richardson number of $Ri_{min} = 1.25$.  The first column illustrates the evolution of the local dust-to-gas ratio $\mu$ (deep red = 0.25, deep blue = 0); the second column shows the evolution of the radial component of vorticity $\omega_x$ (red = vorticity that points into the page, blue = vorticity that points out of the page).  The times corresponding to each frame, in units of the orbital period, are:  3.2, 3.8, 4.5, 5.1, 7.6. }
\end{figure}

We also simulate a thick dust layer that would have been unambiguously stable if there was no Coriolis force.  Figure~\ref{F:run30} shows the two-dimensional nonlinear evolution for the case: $\Sigma_d/\Sigma_g = 0.01$, $\eta_0V_{K0}/c_{s0} = 0.1$, $H_d/H_g = 0.04$, $\mu^{\dagger}_0 = 0.25$ and $Ri_{min} = 1.25$.  The times corresponding to each frame, in units of the orbital period, are: 3.2, 3.8, 4.5, 5.1, 7.6.  The instability sets in at a very small wavelength, fully consistent with the value determined in the linear stability analysis in Figure~\ref{F:growth_coriolis}.  In the next section, we explore the effect of horizontal shear on the evolution of the instability of such thick layers.

\section{3D SIMULATIONS WITH RADIAL SHEAR}

We present a series of fully three-dimensional simulations of settled dust layers with the Coriolis force and differential rotation; Table~\ref{T:simulations} contains a listing of parameters for these simulations.  Dust layers in equilibrium were initialized according to \eqref{E:equilibrium} and \eqref{E:gaussian_dust}, and then perturbations were added to the dust-to gas ratio $\mu$:
\begin{equation}
\mu(x,y,z) =  \mu^{\dagger}(z)\left\{1+A(x,y)\left[\cos(\pi z/2H_{\mu})+\sin(\pi z/2H_{\mu})\right]\right\}.
\end{equation}
The amplitude function $A(x,y)$ is constructed in wavenumber space so that each Fourier mode has random phase and an amplitude inversely proportional to horizontal wavenumber: $ \hat{A}(k_{\perp})\propto k_{\perp}^{-1}$.  The perturbations were also forced to be antisymmetric about the $x$ axis so that the initial kinetic and potential energies~\eqref{E:energy} were unchanged.  In Table~\ref{T:simulations}, the amplitude $A_{rms}$ is the root-mean-square (rms) of these dust-to-gas ratio perturbations in the midplane $z\!=\!0$. 

\begin{table}
\begin{center}
\begin{tabular}{ccccccccc} \tableline
Run & $(L_x, L_y, L_z)$ & $(N_x, N_y, N_z)$ & $\Sigma_d/\Sigma_g$ & $\eta_0 V_{K0}/c_{s0}$ & $H_d/H_g$ & Initial $Ri_{min}$ & $A_{rms}$ & Final $Ri_{min}$\\
\tableline
01 & $(0.1,0.4,0.4)$ & $(32,128,256)$ & 0.01 & 0.1 & 0.01 & 0.0675 & 0.01 & 0.42 \\
02 & $(0.1,0.4,0.4)$ & $(32,128,256)$ & 0.01 & 0.1 & 0.01 & 0.0675 & 0.004 & Stable \\
03 & $(0.1,0.4,0.4)$ & $(32,128,256)$ & 0.01 & 0.1 & 0.01 & 0.0675 & 0.001 & Stable \\
04 & $(0.1,0.4,0.4)$ & $(32,128,256)$ & 0.01 & 0.1 & 0.02 & 0.270 & 0.1 & Stable \\
05 & $(0.1,0.4,0.4)$ & $(32,128,256)$ & 0.01 & 0.1 & 0.02 & 0.270 & 0.2 & Stable \\
06 & $(0.1,0.4,0.4)$ & $(32,128,256)$ & 0.01 & 0.1 & 0.015 & 0.152 & 0.1 & 0.23 \\
07 & $(0.1,0.4,0.4)$ & $(32,128,256)$ & 0.01 & 0.1 & 0.015 & 0.152 & 0.04 & Stable \\
08 & $(0.05,0.2,0.2)$ & $(32,128,256)$ & 0.01 & 0.1 & 0.005 & 0.0169 & $10^{-6}$ & 0.28 \\
09 & $(0.1,0.4,0.4)$ & $(32,128,256)$ & 0.01 & 0.1 & 0.005 & 0.0169 & $10^{-6}$ & 0.38 \\
10 & $(0.1,0.4,0.4)$ & $(32,128,256)$ & 0.02 & 0.1 & 0.01 & 0.0675 & 0.1 & 0.36 \\
11 & $(0.1,0.4,0.4)$ & $(32,128,256)$ & 0.04 & 0.1 & 0.01 & 0.0675 & 0.1 & 0.43 \\
12 & $(0.1,0.4,0.4)$ & $(32,128,256)$ & 0.08 & 0.1 & 0.01 & 0.0675 & 0.1 & 0.47 \\
13 & $(0.1,0.4,0.4)$ & $(32,128,256)$ & 0.02 & 0.1 & 0.02 & 0.270 & 0.1 & Stable \\
14 & $(0.1,0.4,0.4)$ & $(32,128,256)$ & 0.04 & 0.1 & 0.02 & 0.270 & 0.1 & Stable \\
15 & $(0.1,0.4,0.4)$ & $(32,128,256)$ & 0.08 & 0.1 & 0.02 & 0.270 & 0.1 & Stable \\
16 & $(0.05,0.2,0.2)$ & $(32,128,256)$ & 0.02 & 0.1 & 0.005 & 0.0169 & 0.001 & 0.28 \\
17 & $(0.05,0.2,0.2)$ & $(32,128,256)$ & 0.04 & 0.1 & 0.005 & 0.0169 & 0.001 & 0.30 \\
18 & $(0.05,0.2,0.2)$ & $(32,128,256)$ & 0.08 & 0.1 & 0.005 & 0.0169 & 0.001 & 0.40 \\
19 & $(0.1,0.4,0.4)$ & $(32,128,256)$ & 0.01 & 0.2 & 0.02 & 0.0675 & 0.01 & 0.21 \\
20 & $(0.1,0.4,0.4)$ & $(32,128,256)$ & 0.01 & 0.2 & 0.04 & 0.313 & 0.1 & Stable \\
21 & $(0.1,0.4,0.4)$ & $(32,128,256)$ & 0.01 & 0.2 & 0.01 & 0.0169 & 0.01 & 0.22 \\
22 & $(0.05,0.2,0.2)$ & $(32,128,256)$ & 0.01 & 0.05 & 0.01 & 0.270 & 0.04 & Stable \\
23 & $(0.05,0.2,0.2)$ & $(32,128,256)$ & 0.01 & 0.05 & 0.005 & 0.0675 & 0.04 & 0.32 \\
24 & $(0.05,0.2,0.2)$ & $(32,128,256)$ & 0.01 & 0.05 & 0.0025 & 0.0169 & 0.04 & 0.32 \\
25 & $(0.1,0.4,0.4)$ & $(64,256,512)$ & 0.01 & 0.1 & 0.01 & 0.0675 & 0.01 & 0.22 \\
26 & $(0.1,0.4,0.4)$ & $(64,256,512)$ & 0.01 & 0.1 & 0.01 & 0.0675 & 0.001 & Stable \\
27 & $(0.05,0.2,0.2)$ & $(64,256,512)$ & 0.01 & 0.1 & 0.005 & 0.0169 & $10^{-6}$ & 0.21 \\
\tableline
\end{tabular}
\caption{\label{T:simulations} 3D Simulations of settled dust layers.}
\end{center}
\end{table}

\subsection{The dependence on initial amplitude of perturbations}

\begin{figure}
\myplottwo{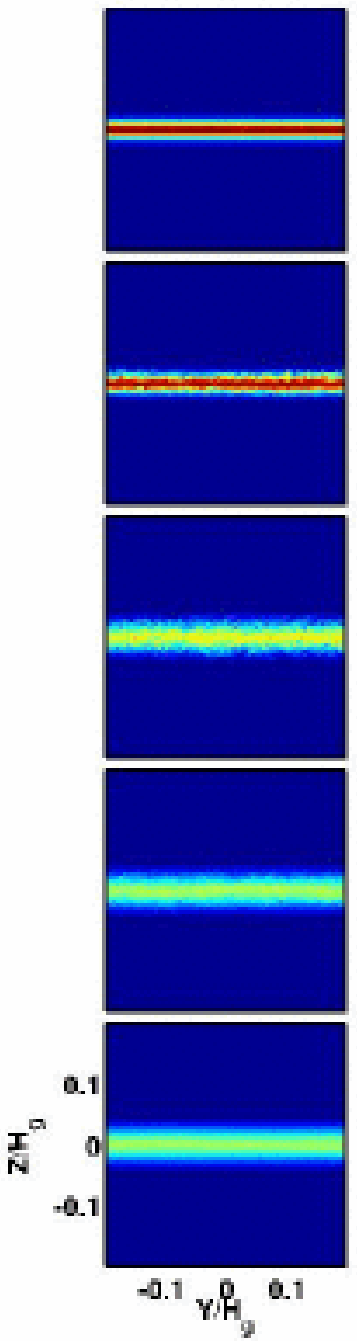}{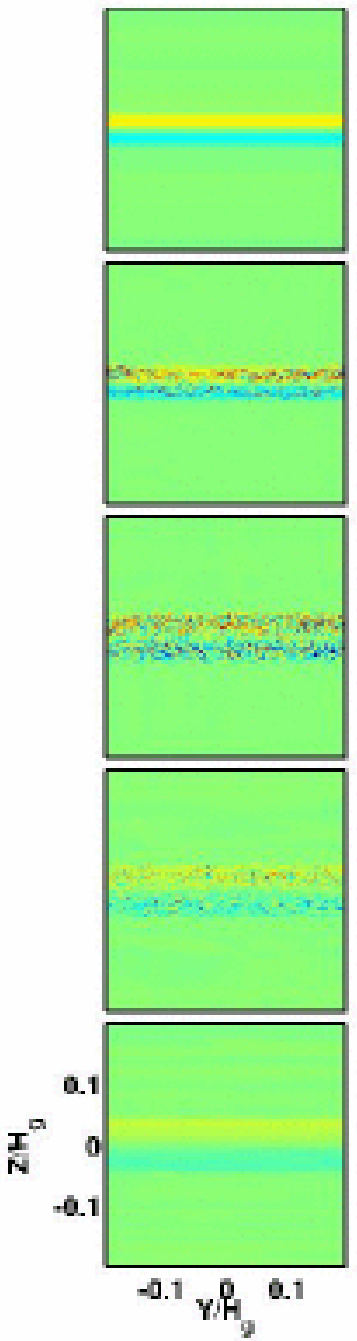}
\caption{\label{F:run25} 3D nonlinear evolution of Kelvin-Helmholtz instability with Coriolis force and horizontal shear. In this simulation, $\Sigma_d/\Sigma_g = 0.01$ and $\eta_0V_{K0}/c_{s0} = 0.1$.  The initial dust scale height is $H_d/H_g = 0.01$, corresponding to a peak local dust-to-gas ratio in the midplane of $\mu^{\dagger}_0 = 1.0$ and a minimum Richardson number of $Ri_{min} = 0.0675$.  The amplitude of initial perturbations was $A_{rms}=10^{-2}$.  The first column illustrates the evolution of the local dust-to-gas ratio $\mu$ (deep red = 1.0, deep blue = 0); the second column shows the evolution of the radial component of vorticity $\omega_x$ (red = vorticity that points into the page, blue = vorticity that points out of the page).  The time interval between frames is 3.4 orbital periods.}
\end{figure}

Figure~\ref{F:run25} shows the results of Run 25, with $\Sigma_d/\Sigma_g = 0.01$, $\eta_0V_{K0}/c_{s0} = 0.1$, $H_d/H_g = 0.01$, $\mu^{\dagger}_0 = 1$, and initial $Ri_{min} = 0.0675$.  These are the same parameters as those used in the 2D runs with no horizontal shear shown in Figure~\ref{F:run28} (no Coriolis force) and Figure~\ref{F:run29} (with Coriolis force).  The perturbations in Run 25 had initial amplitude $A_{rms} = 0.01$ (see also Run 01 at a lower resolution).  Run 26 had exactly the same parameters as Run 25, except the amplitude of perturbations was reduce by a factor of 10 (see also Runs 02 and 03).  This layer is unstable according to the classical Richardson criterion, as demonstrated in the previous section for the cases without horizontal shear.  However, with the addition of horizontal shear, the stability of the dust layer depends also on the amplitude of the initial perturbations: if the magnitude of perturbations is below some threshold, the layer remains stable; whereas if the amplitude exceeds some critical amount, the layer suffers Kelvin-Helmholtz instability.  As  found by \citet{ishitsu03}, unstable eigenmodes have a finite period of growth before the shear stretches them out to high wavenumber and damps further growth.  The nonlinear evolution of a dust layer depends on whether the unstable eigenmodes were able to reach a sufficient amplitude to trigger nonlinear interactions, resulting eventually in turbulence and mixing of the dust with the gas.  We have found through experimentation that the exact critical amplitude depends on such factors as the kind and shape of perturbations (\eg perturbations to dust-to-gas ratio, or temperature, or velocity field), the resolution (number of spectral modes), and the kind and magnitude of small-scale dissipation (\ie hyperviscosity) in the code.   

In Runs 06 and 07, we make the dust layer 50\% thicker than the layer in Run 25: $\Sigma_d/\Sigma_g = 0.01$, $\eta_0V_{K0}/c_{s0} = 0.1$, $H_d/H_g = 0.015$, $\mu^{\dagger}_0 = 0.667$, and initial $Ri_{min} = 0.152$.  The amplitudes of initial perturbations were $A_{rms}=0.1$ and $A_{rms}=0.04$, respectively.  A layer of this thickness would be unstable according to the Richardson criterion.  We again find that the stability depends on the amplitude of perturbations.  Because this layer is closer to stability than the one in Run 25, the eigenmodes would have a slower rate of growth; we would expect that such modes would have to start out at a larger amplitude in order for them to grow to sufficient amplitude to trigger nonlinear effects before the shear damped further growth. This is indeed the case as the critical amplitude is roughly an order of magnitude higher than the for the layer half as thick.

Runs 04 and 05 are for layers that are twice as thick as in Run 25: $\Sigma_d/\Sigma_g = 0.01$, $\eta_0V_{K0}/c_{s0} = 0.1$, $H_d/H_g = 0.02$, $\mu^{\dagger}_0 = 0.5$, and $Ri_{min} = 0.270$.  The amplitudes of initial perturbations were $A_{rms}=0.1$ and $A_{rms}=0.2$, respectively.  These layers are close to the critical Richardson number for stability in the absence of Coriolis force or shear, but would be unstable with the Coriolis force and no horizontal shear.  In 3D simulations with horizontal shear, these layers are found to be stable to even relatively large amplitude perturbations. 
Thus, it appears that the high-Richardson-number unstable flows with the Coriolis force first investigated by \citet{gomez05} are stabilized by the horizontal shear.

\begin{figure}
\myplottwo{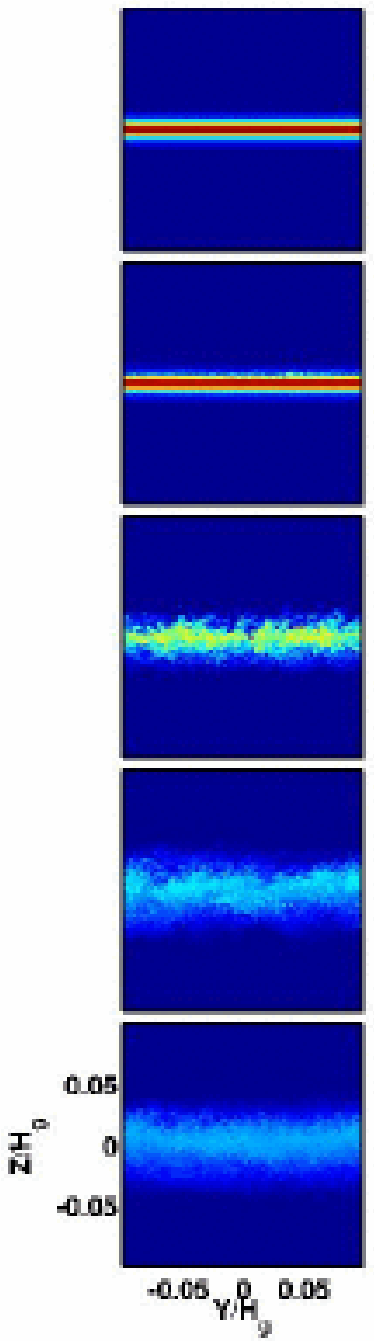}{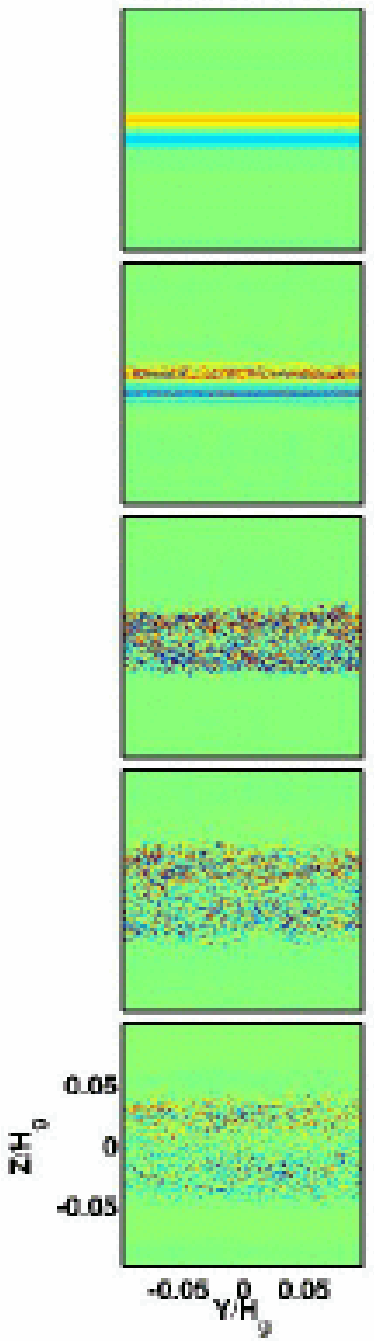}
\caption{\label{F:run27} 3D nonlinear evolution of Kelvin-Helmholtz instability with Coriolis force and horizontal shear. In this simulation, $\Sigma_d/\Sigma_g = 0.01$ and $\eta_0V_{K0}/c_{s0} = 0.1$.  The initial dust scale height is $H_d/H_g = 0.005$, corresponding to a peak local dust-to-gas ratio in the midplane of $\mu^{\dagger}_0 = 2.0$ and a minimum Richardson number of $Ri_{min} = 0.0169$.  The amplitude of initial perturbations was $A_{rms}=10^{-6}$.  The first column illustrates the evolution of the local dust-to-gas ratio $\mu$ (deep red = 2.0, deep blue = 0); the second column shows the evolution of the radial component of vorticity $\omega_x$ (red = vorticity that points into the page, blue = vorticity that points out of the page).  The time interval between frames is 3.4 orbital periods. }
\end{figure}

For thinner layers than those in Run 25, however, the amplitude threshold practically vanishes.  Runs 08, 09, and 27 are for a layer initially half as thick (Richardson number four times smaller): $\Sigma_d/\Sigma_g = 0.01$, $\eta_0V_{K0}/c_{s0} = 0.1$, $H_d/H_g = 0.005$, $\mu^{\dagger}_0 = 2.0$, and $Ri_{min} = 0.0169$ .  The amplitude of perturbations was only $10^{-6}$, yet the layers were still unstable.  Figure~\ref{F:run27} shows the time evolution of this layer.  There may indeed be a threshold, but it would be so low as to be practically irrelevant to the evolution of dust layers in real protoplanetary disk environments.

\subsection{Non-linear mixing and final states}

\begin{figure}
\epsscale{1.0}
\plottwo{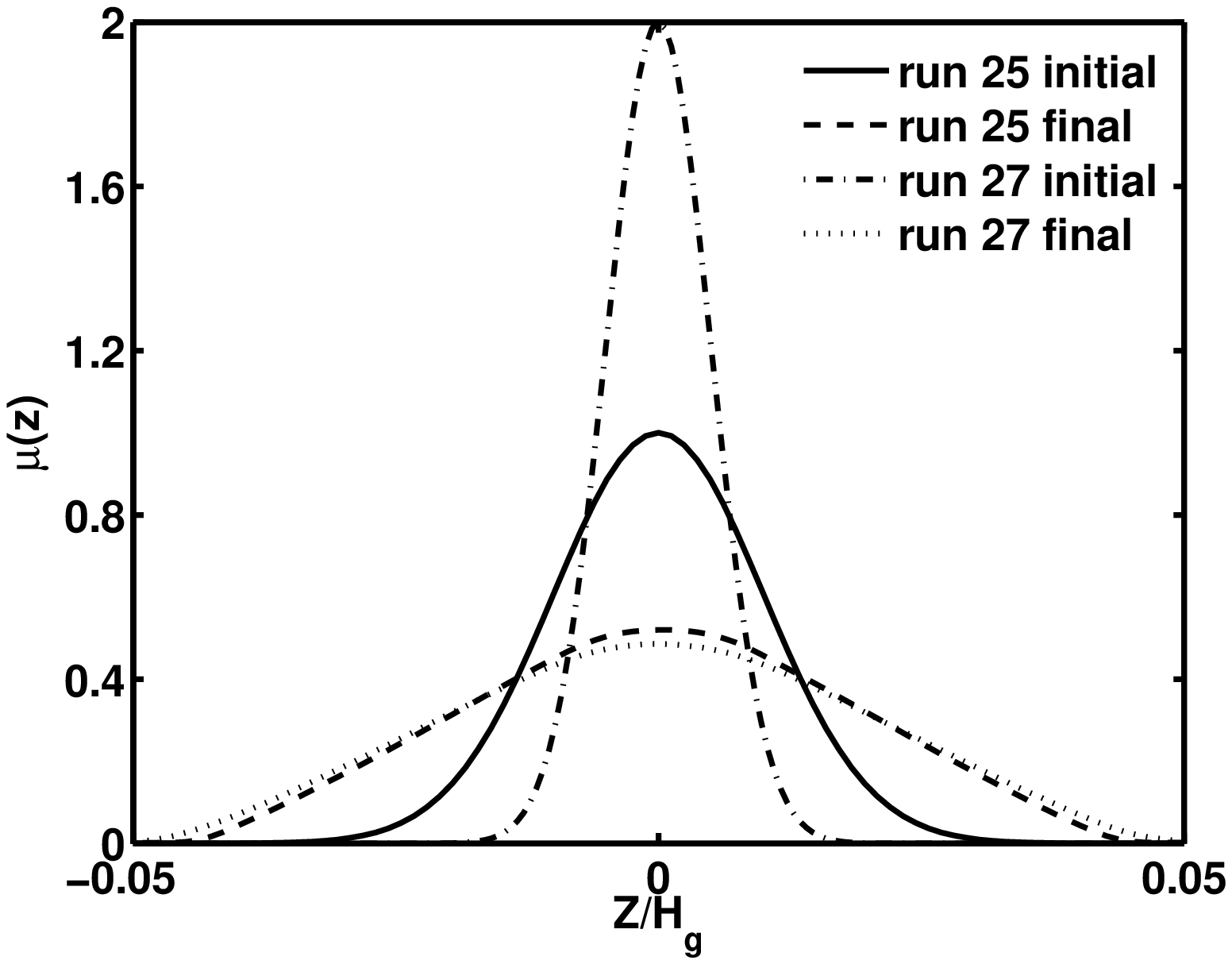}{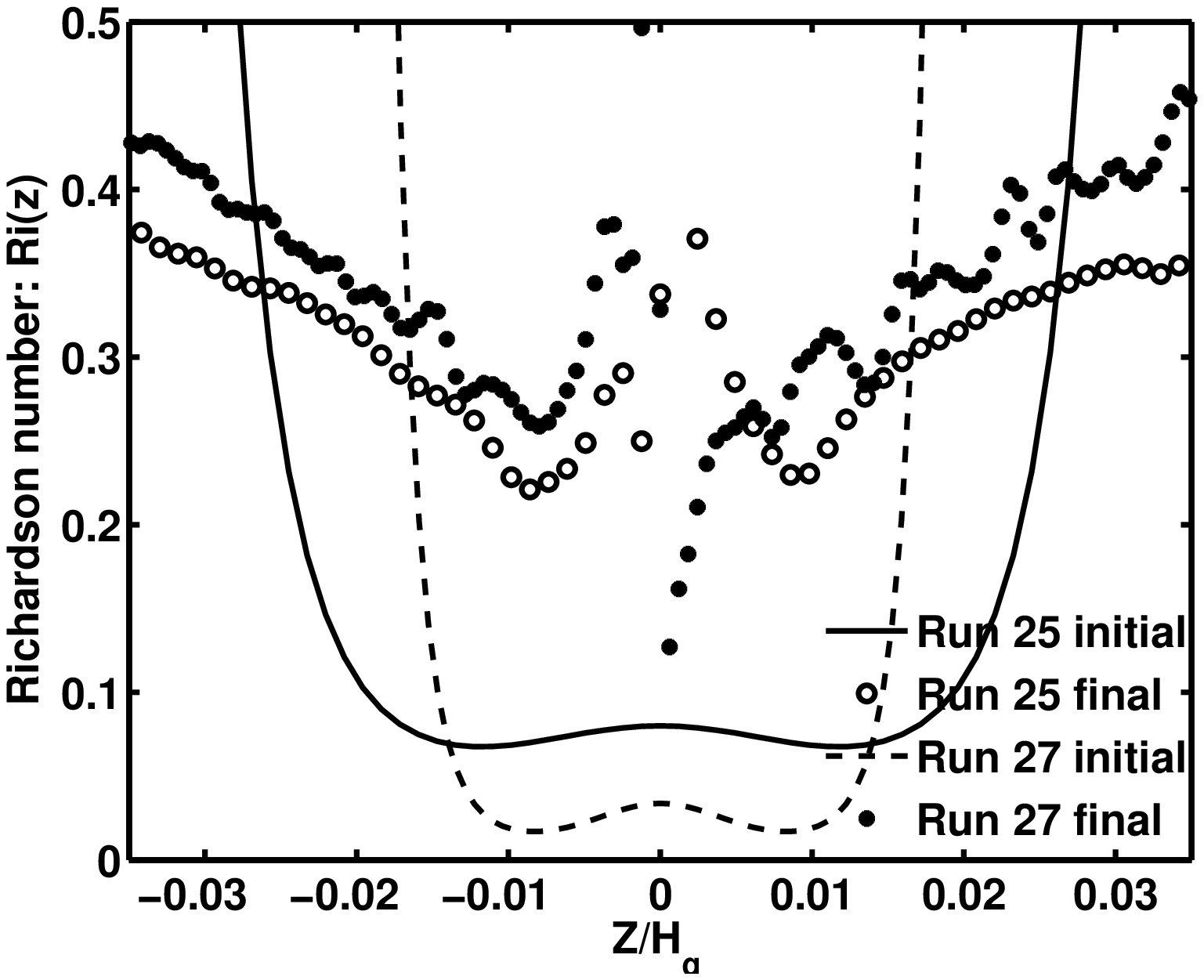}
\caption{\label{F:mu_and_ri} Horizontally-averaged dust-to-gas ratio $\mu$ and Richardson number $Ri$ as a function of height $z$ for initial and final dust layers for runs 25 and 27.}
\end{figure}

\begin{figure}
\epsscale{1.0}
\plotone{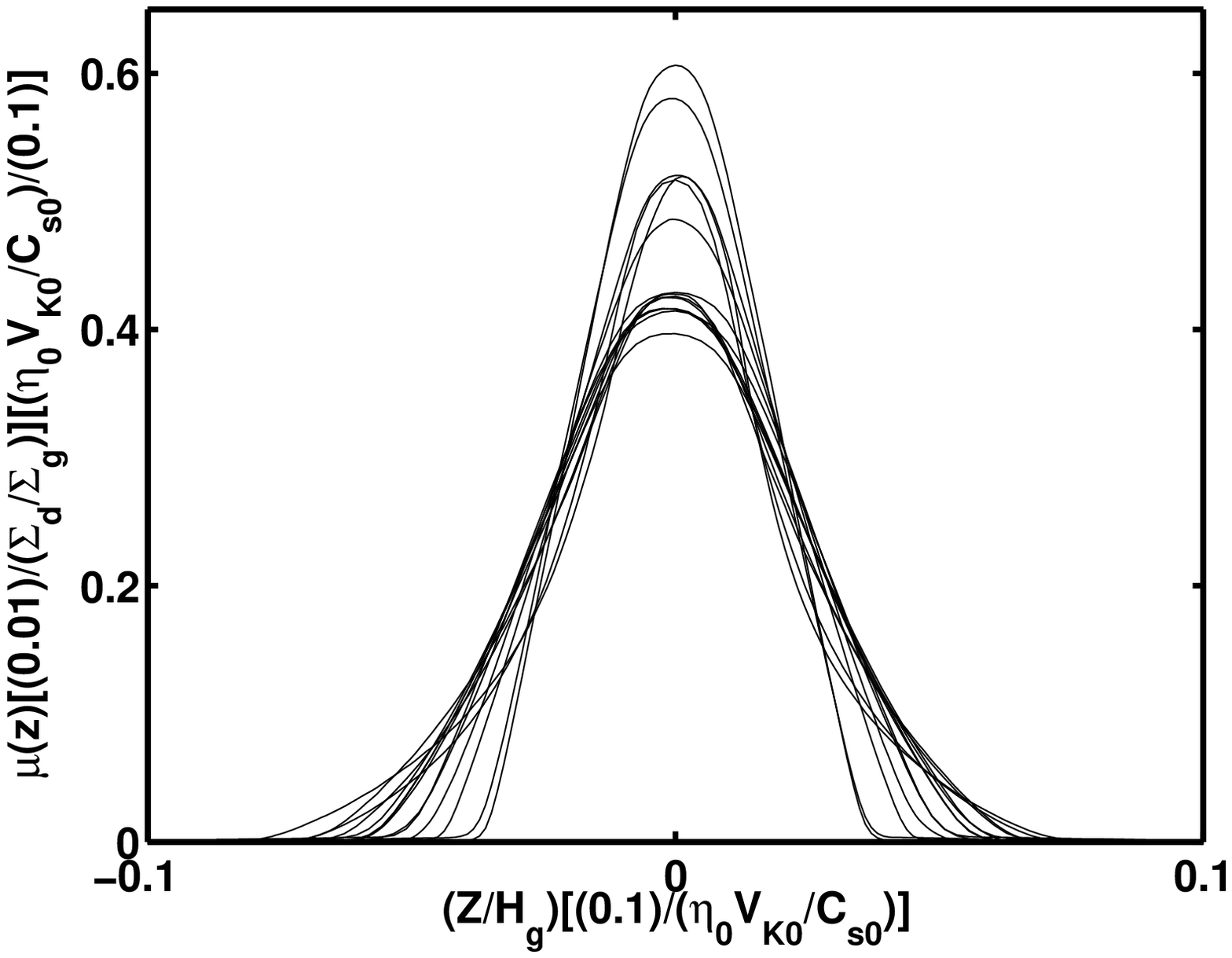}
\caption{\label{F:dtg_eta} Horizontally-averaged dust-to-gas ratio as a function of height for runs with unstable dust layers in Table~\ref{T:simulations}.  Axes are scaled in such a way so that layers that have the same Richardson number would lay on top of one another. }
\end{figure}

In simulations without horizontal shear, the turbulence filled the computational domain and almost completely mixed the dust with the gas.  There was little evidence of a remaining dust layer and the dust-to-gas ratio was nearly uniform.  However, in 3D simulations with horizontal shear, the turbulence and mixing were locally confined to a region above and below the original layer, resulting in the formation of a new, thicker layer.    Figure~\ref{F:mu_and_ri}a  shows the horizontally-averaged dust-to-gas ratio as a function of height for the initial and final dust layers in Runs 25 and 27.  It is interesting to note that the final layers have very nearly the same profile even though they started out with different initial widths with very different growth rates.  In Figure~\ref{F:mu_and_ri}b, the Richardson number as a function of height is graphed for these same two runs.  Because the Richardson number involves the ratio of derivatives that both vanish at the midplane, it is not clear that the few points that have very low Richardson number near the midplane are not just numerical outliers.  If we ignore those few points, then it appears that both of these runs result in minimum Richardson numbers right around the canonical value of one-quarter.  The final column of Table~\ref{T:simulations} shows the numerically computed minimum Richardson number for the cases where the dust layer was unstable.  The lower resolution runs usually result in layers with minimum Richardson numbers that are slightly larger, in the 0.35 -- 0.40 range.  If the resulting turbulence yields a new layer that is thicker than the critical thickness, the dust has no way to re-sediment because these simulations are in the limit of perfect dust-to-gas coupling.  In Figure~\ref{F:dtg_eta}, the final horizontally-averaged dust-to-gas ratio for all the runs with unstable dust layers is plotted as a function of height.  The axes are scaled in such a way so that profiles with the same minimum Richardson number coincide.  The runs included in this figure include simulations with different global dust-to-gas ratios (Runs 10--18), different global gas radial pressure gradients (Runs 19--24), different layer widths, and different amounts of initial perturbations.  Surprisingly, the vast majority of these simulations resulted in final dust layers with nearly the same minimum Richardson number.  

\section{DISCUSSION \& FUTURE WORK}

We revisited the case of no horizontal shear, both with and without the Coriolis force.  The case without the Coriolis force is the most similar to classic Kelvin-Helmholtz instability, and the critical Richardson number for the onset of instability is close to the expected value of one-quarter, consistent with there being sufficient kinetic energy in the shear to lift the heavier fluid out of the gravitational well and mix it with the overlaying lighter fluid \citep{garaud04}.  As first noted by \citet{gomez05}, the case with the Coriolis force is surprisingly different, with instability occurring at much higher Richardson numbers.  However, the wavelengths of the most unstable eigenmodes for these thicker layers is smaller than the thickness of the layer itself.  \citet{ishitsu03} showed that the horizontal shear is able to eventually stabilize unstable eigenmodes by shearing them out to high wavenumber.  However, unstable eigenmodes are able to grow for a period of time; the question is whether they can grow to a large enough amplitude to trigger nonlinear effects and disrupt the dust layer before they are damped as they are sheared.  

In order to investigate the competing influences of the Coriolis force (destabilizing) and horizontal shear (stabilizing), we used a 3D spectral, anelastic, shearing-box code \citep{barranco06} to simulate settled dust layers in the limit of perfect dust-gas coupling.  We find that the stability of dust layers depends on the amplitude of initial perturbations: small perturbations grow for a period of time, but are damped before they reach sufficient magnitude to trigger nonlinear effects; whereas larger amplitude perturbations are able to grow to magnitudes that can disrupt the dust layer, resulting in turbulence and mixing.  Kelvin-Helmholtz instability in thicker layers has slower growth rates, implying that the magnitude of initial amplitudes would have to be larger.  This was seen in Runs 06 and 07 in which the critical threshold for perturbations was an order of magnitude larger for a layer that was 50\% thicker.  Thick dust layers that are stable according to the classic Richardson number criterion, but are unstable with the addition of the Coriolis force exhibit Kelvin-Helmholtz instability with very slow growth rates and at high spatial wavenumber (see Figure~\ref{F:growth_coriolis}).  Three-dimensional simulations of these layers indicate that the horizontal shear is able to damp the instability no matter how large the initial perturbations for these layers (see Runs 04 and 05).  Thinner layers, on the other hand, have very fast growth rates, and the threshold amplitude is so low as to be practically irrelevant in real protoplanetary disks where there is no doubt fluctuations of such low magnitude.

In two-dimensional simulations of unstable dust layers, the turbulence that develops fills the computational domain and leads to large-scale mixing.  Characteristic of ``inverse cascades'' in two-dimensional turbulence, small eddies merge with other eddies to form larger coherent vortices which chaotically advect the dust, homogenizing the dust-to-gas ratio.  In three-dimensional simulations, the turbulence is locally confined to a region right around the original unstable dust layer and does not propagate through the rest of the computational domain.  The unstable thin layers evolve to thicker, more stable layers with minimum Richardson numbers tantalizingly close to the value of one-quarter for the onset of instability in the absence of the Coriolis force and horizontal shear.  These results hold when the global dust to gas ratio $\Sigma_d/\Sigma_g$ and the global gas radial pressure gradient $\eta_0V_{K0}/c_{s0}$ are varied.  There is no reason to expect that the unstable layers in these simulations would evolve to a final state that just happens to be at the critical thickness for stability.  One could imagine that the turbulence is so efficient at mixing that it results in layers whose widths are significantly thicker than thinnest stable layer.  Because the simulations presented here are in the limit of perfect dust-to-gas coupling, no further sedimentation of the dust is allowed so that the thicker final states cannot settle to the critical state.  

Future work will involve two significant improvements.  First, we will relax the perfect dust-to-gas coupling assumption and allow there to be a finite value for the stopping time.  Dust will be treated as a second fluid with its own velocity field \citep{cuzzi93}.  Layers that are unstable will develop turbulence and re-mix the dust with the gas, but then would be allowed to further settle.  \citet{johansen06} performed 2D simulations with two fluids and showed that layers evolved toward a self-regulated state in which further settling was inhibited by turbulence generated by Kelvin-Helmholtz instability, maintaining the layer in a dynamic equilibrium right at the critical thickness for instability.  However, because their simulations were 2D, they found the dust layers had very high Richardson number similar to the layers investigated by \citet{gomez05}.  Second, we will add the effects of non-ideal magnetohydrodynamics with finite resistance.   \citet{turner07} investigated the magneto-rotational instability in protoplanetary disks with ``dead zones'' in the midplane where the ionization is too low to couple the fluid to the magnetic fields.  However, they find that turbulence originating in the cosmic-ray-ionized surface layers can mix free charges into the interior and weakly couple the midplane gas to the magnetic fields, effectively eliminating the dead zone.  Their analysis does not include the role that dust grains play in removing free charges and reducing the ionization.  If turbulence lofts the particles throughout the disk, ionization can be suppressed, which will have the tendency to decouple the interior gas from the magnetic fields, allowing the dead zone to reform.  Dust particle could then re-settle, ionization could increase, re-coupling the gas to the fields and generating a new phase of turbulence.  We plan to investigate if there is indeed a limit cycle to the formation and destruction of the dead zone in protoplanetary disks.

The author would like to thank the National Science Foundation for support via the Astronomy \& Astrophysics Postdoctoral Fellowship program.  Computations for this project were done at
the Institute for Theory \& Computation at the Harvard-Smithsonian Center for Astrophysics, and at the San Diego Supercomputing Center.  The author would like to thank Philip Marcus, Ramesh Narayan, and Niyash Ashfordi for fruitful conversations related to this project.

\bibliographystyle{apj}

\begin{thebibliography}{46}
\expandafter\ifx\csname natexlab\endcsname\relax\def\natexlab#1{#1}\fi

\bibitem[{{Adachi} {et~al.}(1976){Adachi}, {Hayashi}, \& {Nakazawa}}]{adachi76}
{Adachi}, I., {Hayashi}, C., \& {Nakazawa}, K. 1976, Progress of Theoretical
  Physics, 56, 1756

\bibitem[{Barranco \& Marcus(2000)}]{barranco00b}
Barranco, J. \& Marcus, P. 2000, in Center for {T}urbulence {R}esearch --
  {P}roceedings of the 2000 {S}ummer {P}rogram, 97--108

\bibitem[{Barranco \& Marcus(2005)}]{barranco05}
Barranco, J. \& Marcus, P. 2005, \apj, 623, 1157

\bibitem[{Barranco \& Marcus(2006)}]{barranco06}
---. 2006, J. Comp. Phys., 219, 21

\bibitem[{Barranco {et~al.}(2000)Barranco, Marcus, \& Umurhan}]{barranco00a}
Barranco, J., Marcus, P., \& Umurhan, M. 2000, in Center for {T}urbulence
  {R}esearch -- {P}roceedings of the 2000 {S}ummer {P}rogram, 85--96

\bibitem[{Beckwith {et~al.}(2000)Beckwith, Henning, \& Nakagawa}]{beckwith00}
Beckwith, S., Henning, T., \& Nakagawa, Y. 2000, in Protostars and Planets IV,
  ed. V.~Mannings, A.~Boss, \& S.~Russell (Tuscon: University of Arizona
  Press), 533--558

\bibitem[{Boyd(1989)}]{boyd89}
Boyd, J. 1989, Chebyshev and Fourier Spectral Methods (New York:
  Springer-Verlag)

\bibitem[{{Brittain} {et~al.}(2005){Brittain}, {Rettig}, {Simon}, \&
  {Kulesa}}]{brittain05}
{Brittain}, S.~D., {Rettig}, T.~W., {Simon}, T., \& {Kulesa}, C. 2005, \apj,
  626, 283

\bibitem[{Cain {et~al.}(1984)Cain, Ferziger, \& Reynolds}]{cain84}
Cain, A., Ferziger, J., \& Reynolds, W. 1984, J. Comp. Phys., 56, 272

\bibitem[{Canuto {et~al.}(1988)Canuto, Hussaini, Quarteroni, \&
  Zang}]{canuto88}
Canuto, C., Hussaini, M., Quarteroni, A., \& Zang, T. 1988, Spectral Methods in
  Fluid Dynamics (New York: Springer-Verlag)

\bibitem[{Champney {et~al.}(1995)Champney, Dobrovolskis, \& Cuzzi}]{champney95}
Champney, J., Dobrovolskis, A., \& Cuzzi, J. 1995, Phys. Fluids, 7, 1703

\bibitem[{Chandrasekhar(1960)}]{chandra60}
Chandrasekhar, S. 1960, Proc. Natl. Acad. Sci. USA, 46, 253

\bibitem[{Chandrasekhar(1961)}]{chandra61}
---. 1961, Hydrodynamic and Hydromagnetic Stability (New York: Dover)

\bibitem[{Cuzzi {et~al.}(1993)Cuzzi, Dobrovolskis, \& Champney}]{cuzzi93}
Cuzzi, J., Dobrovolskis, A., \& Champney, J. 1993, Icarus, 106, 102

\bibitem[{{Dobrovolskis} {et~al.}(1999){Dobrovolskis}, {Dacles-Mariani}, \&
  {Cuzzi}}]{dobrovolskis99}
{Dobrovolskis}, A.~R., {Dacles-Mariani}, J.~S., \& {Cuzzi}, J.~N. 1999, \jgr,
  104, 30805

\bibitem[{Drazin \& Reid(1981)}]{drazin81}
Drazin, P. \& Reid, W. 1981, Hydrodynamic Stability (Cambridge: Cambridge
  University Press)

\bibitem[{{Dullemond} {et~al.}(2007){Dullemond}, {Henning}, {Visser}, {Geers},
  {van Dishoeck}, \& {Pontoppidan}}]{dullemond07}
{Dullemond}, C.~P., {Henning}, T., {Visser}, R., {Geers}, V.~C., {van
  Dishoeck}, E.~F., \& {Pontoppidan}, K.~M. 2007, \aap, 473, 457

\bibitem[{Frank {et~al.}(1985)Frank, King, \& Raine}]{frank85}
Frank, J., King, A., \& Raine, D. 1985, Accretion Power in Astrophysics
  (Cambridge: Cambridge University Press)

\bibitem[{{Garaud} {et~al.}(2004){Garaud}, {Barri{\`e}re-Fouchet}, \&
  {Lin}}]{garaud04a}
{Garaud}, P., {Barri{\`e}re-Fouchet}, L., \& {Lin}, D.~N.~C. 2004, \apj, 603,
  292

\bibitem[{{Garaud} \& {Lin}(2004)}]{garaud04}
{Garaud}, P. \& {Lin}, D.~N.~C. 2004, \apj, 608, 1050

\bibitem[{Gilman \& Glatzmaier(1981)}]{gilman81}
Gilman, P. \& Glatzmaier, G. 1981, \apjs, 45, 335

\bibitem[{Glatzmaier \& Gilman(1981{\natexlab{a}})}]{glatzmaier81a}
Glatzmaier, G. \& Gilman, P. 1981{\natexlab{a}}, \apjs, 45, 351

\bibitem[{Glatzmaier \& Gilman(1981{\natexlab{b}})}]{glatzmaier81b}
---. 1981{\natexlab{b}}, \apjs, 45, 381

\bibitem[{Goldreich \& Lynden-Bell(1965)}]{goldreich65b}
Goldreich, P. \& Lynden-Bell, D. 1965, \mnras, 130, 125

\bibitem[{Goldreich \& Ward(1973)}]{goldreich73}
Goldreich, P. \& Ward, W. 1973, \apj, 183, 1051

\bibitem[{{G{\'o}mez} \& {Ostriker}(2005)}]{gomez05}
{G{\'o}mez}, G.~C. \& {Ostriker}, E.~C. 2005, \apj, 630, 1093

\bibitem[{Gottlieb \& Orszag(1977)}]{gottlieborszag77}
Gottlieb, D. \& Orszag, S. 1977, Numerical Analysis of Spectral Methods: Theory
  and Applications (Philadelphia: Society for Industrial and Applied
  Mathematics)

\bibitem[{Gough(1969)}]{gough69}
Gough, D. 1969, J. Atmos. Sci., 26, 448

\bibitem[{{Ishitsu} \& {Sekiya}(2003)}]{ishitsu03}
{Ishitsu}, N. \& {Sekiya}, M. 2003, Icarus, 165, 181

\bibitem[{{Johansen} {et~al.}(2006){Johansen}, {Henning}, \&
  {Klahr}}]{johansen06}
{Johansen}, A., {Henning}, T., \& {Klahr}, H. 2006, \apj, 643, 1219

\bibitem[{Lissauer(1993)}]{lissauer93}
Lissauer, J. 1993, \araa, 31, 129

\bibitem[{Marcus(1986)}]{marcus86a}
Marcus, P. 1986, in Proceedings of Astrophysical Radiation Hydrodynamics, ed.
  K.-H. Winkler \& M.~Norman (Springer-Verlag), 359--386

\bibitem[{Marcus \& Press(1977)}]{marcus77}
Marcus, P. \& Press, W. 1977, J. Fluid Mech., 79, 525

\bibitem[{Ogura \& Phillips(1962)}]{ogura62}
Ogura, Y. \& Phillips, N. 1962, J. Atmos. Sci., 19, 73

\bibitem[{{Rettig} {et~al.}(2006){Rettig}, {Brittain}, {Simon}, {Gibb},
  {Balsara}, {Tilley}, \& {Kulesa}}]{rettig06}
{Rettig}, T., {Brittain}, S., {Simon}, T., {Gibb}, E., {Balsara}, D.~S.,
  {Tilley}, D.~A., \& {Kulesa}, C. 2006, \apj, 646, 342

\bibitem[{Rogallo(1981)}]{rogallo81}
Rogallo, R. 1981, Numerical experiments in homogeneous turbulence, Technical
  memorandum 81315, NASA

\bibitem[{Ryu \& Goodman(1992)}]{ryu92}
Ryu, D. \& Goodman, J. 1992, \apj, 388, 438

\bibitem[{Safronov(1969)}]{safronov69}
Safronov, V. 1969, Evolution of the Protoplanetary Cloud and the Formation of
  the Earth and Planets (Moscow: Nauka Press)

\bibitem[{{Sekiya}(1998)}]{sekiya98}
{Sekiya}, M. 1998, Icarus, 133, 298

\bibitem[{{Sekiya} \& {Ishitsu}(2000)}]{sekiya00}
{Sekiya}, M. \& {Ishitsu}, N. 2000, Earth, Planets, and Space, 52, 517

\bibitem[{Squire(1933)}]{squire33}
Squire, H. 1933, Proc. R. Soc. London A, 142, 621

\bibitem[{{Turner} {et~al.}(2007){Turner}, {Sano}, \&
  {Dziourkevitch}}]{turner07}
{Turner}, N.~J., {Sano}, T., \& {Dziourkevitch}, N. 2007, \apj, 659, 729

\bibitem[{Weidenschilling(1977)}]{weidenschilling77}
Weidenschilling, S. 1977, \mnras, 180, 57

\bibitem[{{Weidenschilling}(1980)}]{weidenschilling80}
{Weidenschilling}, S.~J. 1980, Icarus, 44, 172

\bibitem[{{Youdin} \& {Chiang}(2004)}]{youdin04}
{Youdin}, A.~N. \& {Chiang}, E.~I. 2004, \apj, 601, 1109

\bibitem[{{Youdin} \& {Shu}(2002)}]{youdin02}
{Youdin}, A.~N. \& {Shu}, F.~H. 2002, \apj, 580, 494

\end{thebibliography}

\end{document}